\documentclass[12pt]{article}
\usepackage{latexsym,amssymb,epsfig,rotating}

\def\0{\phantom{0}}

\begin{document}
\pagenumbering{arabic}
\baselineskip25pt

\begin{center}
{\bf \large Prediction of ternary vapor-liquid equilibria for 33 systems by molecular simulation} \\

\bigskip
\renewcommand{\thefootnote}{\fnsymbol{footnote}}
Yow-lin Huang$^1$, Jadran Vrabec\footnote[1] 
{corresponding author, tel.: +49-5251/60-2421, 
fax: +49-5251/60-3522, \\ email: jadran.vrabec@upb.de}$^1$, 
Hans Hasse$^2$
\renewcommand{\thefootnote}{\arabic{footnote}} \\
\end{center}

\noindent
$^1$ Lehrstuhl f\"ur Thermodynamik und Energietechnik, Universit\"at Paderborn, Warburger Stra\ss e 100, 33098 Paderborn, Germany \\

\noindent
$^2$ Laboratory for Engineering Thermodynamics, University of Kaiserslautern, Erwin-Schr\"odinger-Stra\ss e 44, 67663 Kaisers\-lautern, Germany


\bigskip
\bigskip
{\bf Abstract}

\noindent
 
A set of molecular models for 78 pure substances from prior work is taken as a basis for systematically studying vapor-liquid equilibria (VLE) in ternary systems.
All 33 ternary mixtures of these 78 components for which experimental VLE data is available are studied by molecular simulation. 
The mixture models are based on the modified Lorentz-Berthelot combining rule that contains one binary interaction parameter which was adjusted 
to a single experimental binary vapor pressure of each binary subsystem in prior work.
No adjustment to ternary data is carried out. 
The predictions from the molecular models of the 33 ternary mixtures are compared to the available experimental data.
In almost all cases, the molecular models give excellent predictions of the ternary mixture properties.

\clearpage

\section{INTRODUCTION}
In previous work of our group, a set of molecular models has been developed for 78 real pure fluids using the dipolar or quadrupolar two-center Lennard-Jones (2CLJD and 2CLJQ) potential \cite{vrabec0112,vrabec0311}. 
This model type has been proposed more than three decades ago \cite{streett7735}, however, it is far from being fully exploited. 
Polar 2CLJ models consider the basic molecular interactions repulsion and dispersive attraction and also feature anisotropy and polarity in a simple way.
78 small molecules consisting of up to nine atoms that belong to different classes of real fluids, including noble gases, alkanes, halogens and numerous refrigerants, were modeled using that approach \cite{vrabec0112,vrabec0311}.
For many of the 78 molecules, the polar 2CLJ model is only a crude assumption. 
E.g., the asymmetry of molecules is neglected and the polar interaction is always aligned along the molecular axis. 
Also the polarizability, which is often assumed to be a crucial molecular property for thermodynamics, is only implicitly considered by Lennard-Jones interaction sites.
Furthermore, the internal degrees of freedom are neglected as the polar 2CLJ models are rigid. 

After showing that these simple molecular models give excellent results for vapor-liquid equilibria (VLE) of both the pure components and their binary mixtures \cite{vrabec0112,vrabec0311,Zbinary}, 
the aim of the present work is to investigate on a broad basis whether such models are fully transferable to VLE of ternary systems. 

Based on the 78 pure substance models \cite{vrabec0112,vrabec0311}, the unlike energy parameter was adjusted in previous work 
\cite{Zbinary,vrabec0321,vrabec0521,stollv9} to the experimental binary vapor pressure for 267 binary systems in order to very accurately describe their VLE. 
The direct transferability of such models to higher systems was also shown by our group \cite{vrabec0321,vrabec0521,stollv9} with VLE predictions of five ternary mixtures.
This work is extended here to 33 ternary systems. 

A few publications on molecular simulation results for ternary VLE are available from different authors: 
Carrero-Mantilla and Llano-Restrepo \cite{carrero} (N$_2$ + CH$_4$ + C$_2$H$_6$), 
Potoff and Siepmann \cite{Potoff2001} (N$_2$ + CO$_2$ + propane), 
Kamath and Potoff \cite{Kamath2006} (CH$_4$ + H$_2$S + CO$_2$), 
Hansen et al. \cite{Hansen2006} (N$_2$ + O$_2$ + CO$_2$),
Liu and Beck \cite{Liu1998} (CH$_4$ + CO$_2$ + C$_2$H$_6$),
Nath et al. \cite{Nath2001} (C$_2$H$_4$ + 1-hexene + polyethylene),
L\'isal et al. \cite{lisal} (isobutene + methanol + MTBE)
and Van't Hof \cite{Arons2005} (CH$_4$ + CO$_2$ + C$_2$H$_6$ and CH$_4$ + CO$_2$ + propane).
However, each of these publications is restricted to one or two ternary mixtures only. 
Note that there are additional works on ternary VLE by simulation \cite{Tsang1995,Miguel1997,Escobedo1999,Sadus1999,Errington2005,Attwood2008}, but they deal exclusively with model systems.

To avoid an arbitrary selection of the studied systems, a combinatorial brute force approach was taken here.
Theoretically, out of the $N=78$ components modeled in \cite{vrabec0112,vrabec0311} $N(N-1)(N-2)/6 = 76~076$ ternary mixtures can be formed, but of course, by far not all of these systems have been studied experimentally.
To our knowledge, VLE were measured only for a subset of 33 out of the 76 076 ternary systems, corresponding to 0.043\%.
In the present work, all these 33 ternary mixtures were studied.
This is the largest set of ternary systems that was used so far to probe the application of molecular modeling and simulation to ternary mixtures.

It would have been attractive to investigate VLE of multi-component mixtures with more than three components too, however, to our knowledge no experimental data exists for any mixture consisting of four or more of those 78 components. 

The simulation results from the present work are compared to experimental data and in most cases to the Peng-Robinson equation of state (EOS) \cite{pengrob} which was applied in the same predictive way, i.e. it was adjusted to the same binary data that was also used to adjust the molecular mixture models, cf. \cite{Zbinary}.
Due to the fact that the Peng-Robinson EOS is widely known, it is not described here, for details see \cite{Zbinary}.

\section{EXPERIMENTAL DATABASE}
\label{expdatabase}
Experimental data were predominately retrieved using Dortmunder Datenbank (DDB) \cite{ddb}, which collects {\it all} publicly available mixture VLE data sets, covering more than a century of experimental work. 
According to DDB, for a subset of 33 of the potential 76 076 ternary mixtures experimental VLE data is available.
They stem from 30 publications \cite{vleN2ArCH4}-\cite{vleR140aR142bR141b}.
These 33 ternary systems include 35 of the 78 pure components; cf. Table \ref{pureall} for the full component list including their CAS RN number for proper identification.
Please note that the ASHRAE nomenclature is preferred in the following due to its brevity, despite its deficiencies \cite{deiters}. 

The studied 33 ternary systems are listed in Table \ref{terall} together with a reference to the experimental VLE data.
Of those 33 ternary mixtures, five have been modeled in previous work of our group \cite{vrabec0321,vrabec0521,stollv9} but the resulting VLE data were published only partly.

It can be argued that these 33 systems, being just 0.043\% of the full combinatorial sample, were selected by the experimentalists due their technical or scientific importance.

\section{PURE FLUID MODELS}
As explained above, 35 polar 2CLJ molecular models, taken from \cite{vrabec0112,vrabec0311}, were used here.
A list of the pure fluids is given in Table \ref{pureall}. 
These are two spherical non-polar (1CLJ) models for Ar and ${\rm CH_4}$, three spherical dipolar (Stockmayer or 1CLJD) models for R30, R30B2 and R32, furthermore 17 elongated dipolar (2CLJD) models which include CO and numerous refrigerants, and finally 13 elongated quadrupolar (2CLJQ) models which include N$_2$, O$_2$, alkanes, refrigerants and ${\rm CO_2}$.

A detailed description of the polar two-center Lennard-Jones pair potential is provided in \cite{Zbinary} and not repeated here. 
Polar 2CLJ models have four parameters: size $\sigma$, energy $\epsilon$, elongation $L$ and either dipolar momentum $\mu$ or quadrupolar momentum $Q$. 
Stockmayer models have a vanishing elongation, while the non-polar spherical LJ models have only $\sigma$ and $\epsilon$. 
Model parameters were adjusted in \cite{vrabec0112,vrabec0311} to experimental pure fluid VLE data using global correlations of critical temperature, saturated liquid density and vapor pressure as functions of these molecular parameters \cite{stoll0133,stoll0329}. 
These pure substance model parameters are also not repeated here.
It should be noted that a wide range of polar momenta are covered by the 35 pure substance models regarded in this work. 
Starting from a non-existent polar momentum in case of Ar and ${\rm CH_4}$, it ranges to up to 3.7104 D for the dipolar R30B2 and up to 16.143 D\AA{} for the quadrupolar R1110. 

The advantage of these molecular models is their simplicity, which reduces simulation time considerably, and their
accuracy: typically, the relative deviations between simulation and experiment are below 1~\% for the saturated liquid density, below 3~\% for the vapor pressure, and below 3~\% for the enthalpy of vaporization 
\cite{vrabec0112,vrabec0311}. 
They also have shown to reliably predict Joule-Thomson inversion curves for pure fluids and mixtures \cite{vrabec0525,Ashish}, covering a wide range of state points, and also transport properties \cite{vrabec0417,vrabec21,vrabec24,vrabec27,gimmynew}.

\section{MOLECULAR MIXTURE MODELS}
\label{mixturemodels}
On the basis of pairwise additive pure fluid potentials, molecular modeling of mixtures reduces to modeling the interactions between unlike molecules. 
Unlike interactions consist of two different types here. 
On the one hand there are the electrostatic interactions (dipole-dipole, dipole-quadrupole, and quadrupole-quadrupole). 
These interactions are treated in a physically straightforward way, simply using the laws of electrostatics \cite{Zbinary}.

Repulsion and dispersive attraction are other interaction types and are present between all molecules.
If a mixture A + B + C is modeled on the basis of Len\-nard-Jones potentials, the knowledge of three pairs of unlike Lennard-Jones parameters is required:
$\sigma_{\rm AB}$, $\epsilon_{\rm AB}$ and $\sigma_{\rm AC}$, $\epsilon_{\rm AC}$ as well as $\sigma_{\rm BC}$, $\epsilon_{\rm BC}$. 
For their determination, the broadly used Lorentz-Berthelot combining rule is a good starting point \cite{Schnabel2006}. 
However, introducing a binary interaction parameter $\xi$ to adjust the unlike energy parameter $\epsilon_{ij}$ 
\begin{equation}
\sigma_{ij} = \left( \sigma_{i}\!+\!\sigma_{j} \right)\!/2 \label{sigLB},
\end{equation}
and 
\begin{equation}
\epsilon_{ij}=\xi \sqrt{\epsilon_{i} \epsilon_{j}}, \label{xiepsLB}
\end{equation}
allows almost always for an optimal representation of the binary fluid phase behavior \cite{Zbinary}.

For VLE, it was shown in \cite{Schnabel2006} that $\xi$ can be adjusted to a single experimental binary vapor pressure.
Values for $\xi$ are given in \cite{Zbinary} for 267 binary combinations. 
Note that the present 33 ternary systems comprise 65 different binary subsystems, whereof 62 were covered in \cite{Zbinary}. 
The three exceptions are N$_2$ + R14, R125 + R161 and R134a + R161.
It was abstained here to adjust the binary interaction parameter for these three binary subsystems to ternary VLE data, thus $\xi=1$ was specified instead. 
We refrained here from adjusting the binary interaction parameter $k_{ij}$ of the Peng-Robinson EOS for these.

\section{RESULTS AND DISCUSSION}
\label{discussionresults}
To assess the predictive quality of the mixture models, ternary VLE were determined by molecular simulation predominantly at state points for which a direct comparison to experimental data is possible. 
Simulation details are given in the Appendix.
The Grand Equilibrium method \cite{vrabec15} was used for the VLE simulations, where temperature and liquid composition are the independently specified thermodynamic variables, while vapor pressure, saturated vapor composition, saturated densities and enthalpy of vaporization are determined. 
In most cases, simulation results are presented that match exactly with the experimental bubble line composition. 
However, if it was found that there is a significant mismatch for the resulting vapor pressure with respect to the experiment, the simulative bubble line composition was altered so that both data sets are almost in the same temperature-pressure plane.

As experimental densities and enthalpies are rarely available in the literature, only vapor pressure and saturated vapor composition were used for this assessment. 
It should be noted that saturated vapor composition data is available for 26 of the investigated 33 ternary systems, for the remaining seven systems, cf. Table \ref{terall}, only bubble line data is available.

\begin{figure}[ht]																																							
\begin{center}							
\caption[Ternary vapor-liquid equilibrium phase diagram of the mixture Ar + ${\rm CH_4}$ + ${\rm N_2}$ at 112 K and 0.91 MPa: {\Large $\bullet$} present simulation data, $+$ experimental data \cite{vleN2ArCH4}, --- Peng-Robinson equation of state.]{}						
\bigskip
\epsfig{file=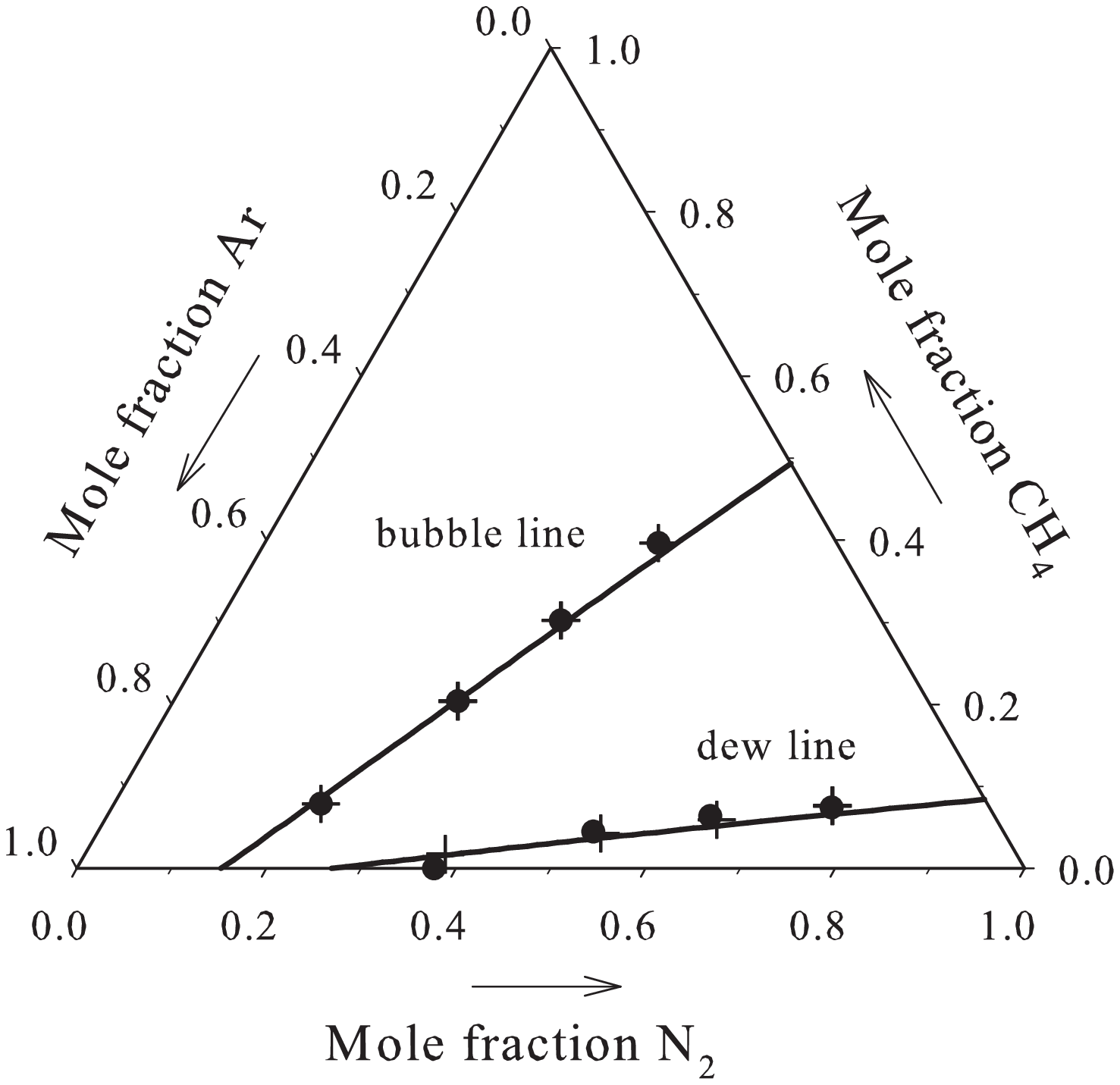,scale=0.56}																												
\label{ar_ch4_n2}
\end{center}																																			
\end{figure}

\begin{figure}[ht]																																							
\begin{center}																																							
\caption[Ternary vapor-liquid equilibrium phase diagram of the mixture Ar + ${\rm CH_4}$ + CO at 164 K and 3.546 MPa: {\Large $\bullet$} present simulation data, $+$ experimental data \cite{vleCOArCH4}, --- Peng-Robinson equation of state.]{}																																							
\bigskip																																							
\epsfig{file=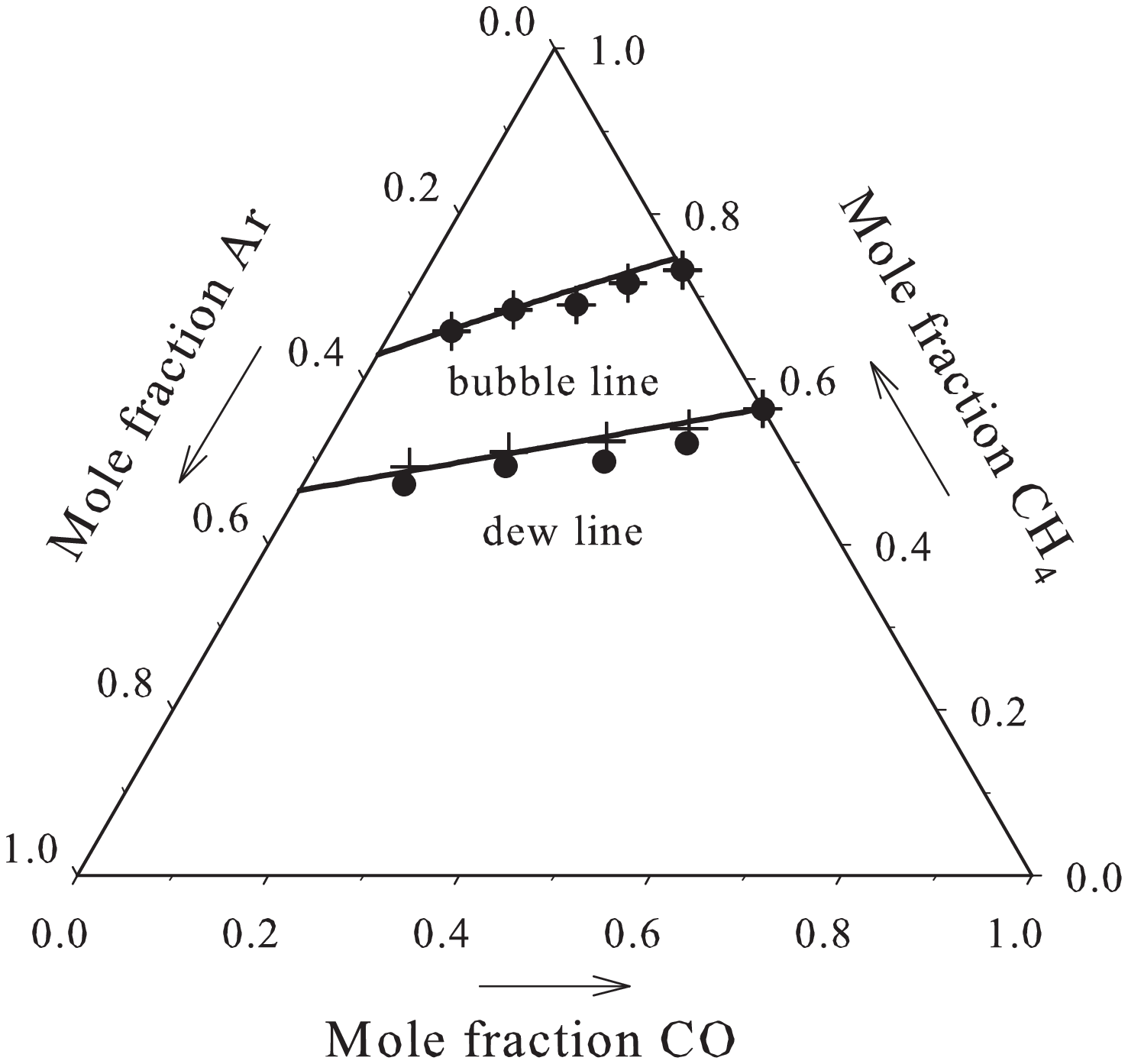,scale=0.56}																																							
\label{ar_ch4_co} 	 																																							
\end{center}																																							
\end{figure}

\begin{figure}[ht]																																							
\begin{center}																																							
\caption[Ternary vapor-liquid equilibrium phase diagram of the mixture Ar + ${\rm CH_4}$ + ${\rm C_2H_6}$ at 115 K and 0.412 MPa: {\Large $\bullet$} present simulation data, $+$ experimental data \cite{vleArCH4C2H6}, --- Peng-Robinson equation of state.]{}																																							
\bigskip																																							
\epsfig{file=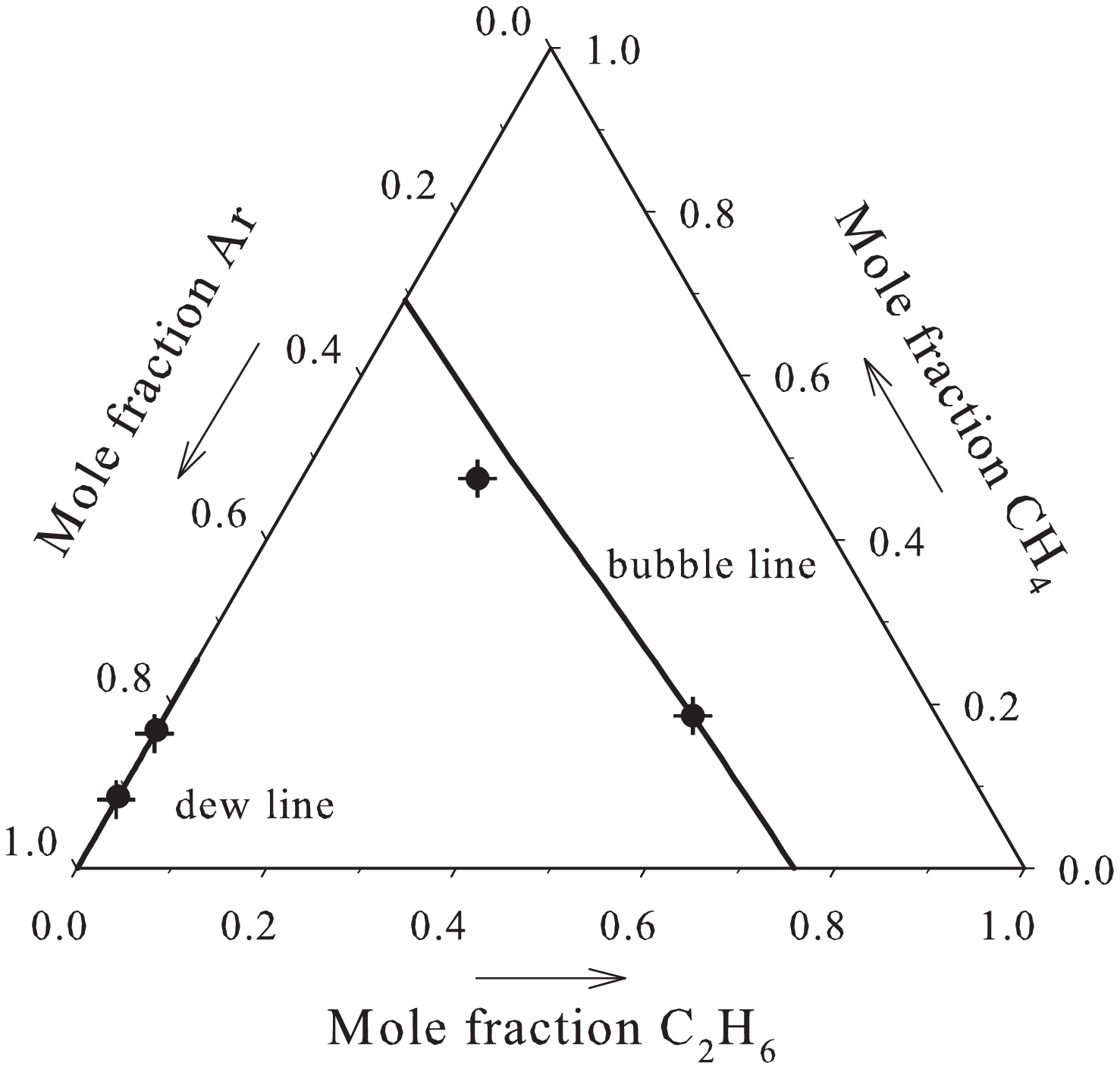,scale=0.56}																																							
\label{ar_ch4_c2h6} 	 																																							
\end{center}																																							
\end{figure}

\begin{figure}[ht]																																		 	 																																							
\begin{center}																																							
\caption[Ternary vapor-liquid equilibrium phase diagram of the mixture Ar + ${\rm N_2}$ + ${\rm O_2}$ at 84.7 K and 0.204 MPa: {\Large $\bullet$} present simulation data, $+$ experimental data \cite{vleN2ArO2}, --- Peng-Robinson equation of state.]{}																																							
\bigskip																																							
\epsfig{file=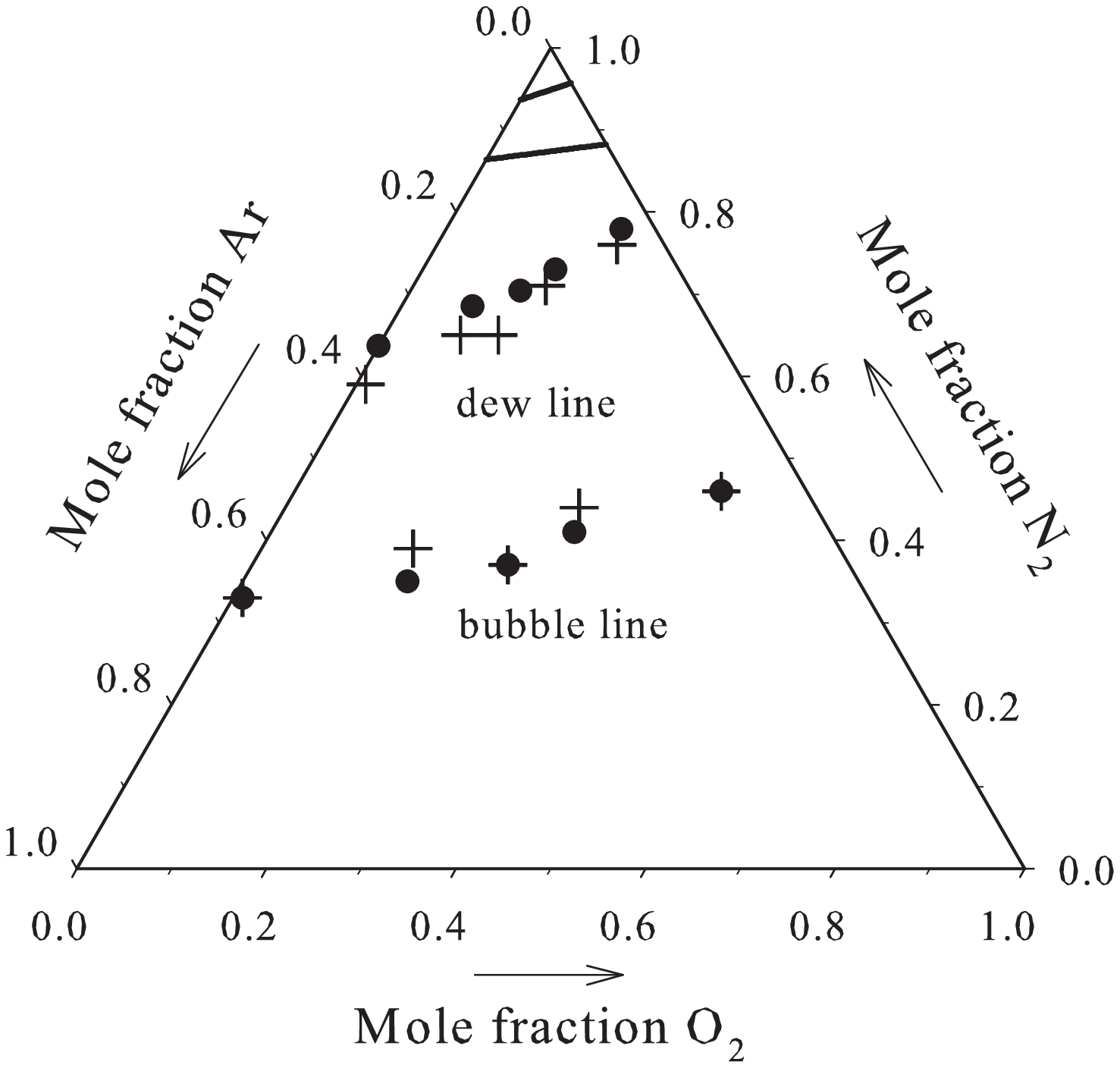,scale=0.56}																																							
\label{ar_n2_o2}
\end{center}																																							
\end{figure}

\begin{figure}[ht]																																							
\begin{center}																																							
\caption[Ternary vapor-liquid equilibrium phase diagram of the mixture ${\rm CH_4}$ + ${\rm N_2}$ + CO at 140 K and 2 MPa: {\Large $\bullet$} present simulation data, $+$ experimental data \cite{vleN2COCH4}, --- Peng-Robinson equation of state.]{}																														
\bigskip																																	
\epsfig{file=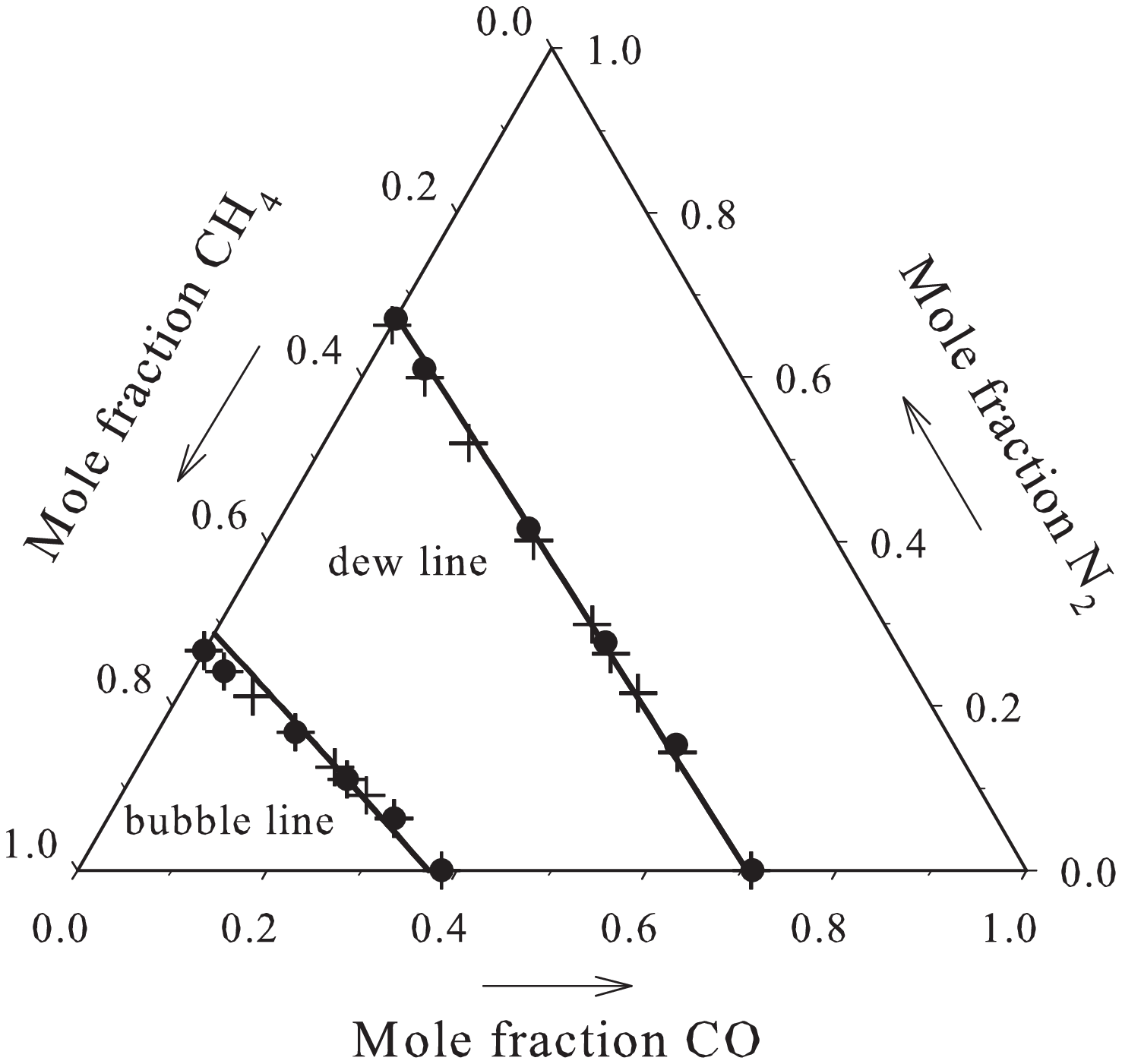,scale=0.56}																																							
\label{ch4_n2_o2} 	 																																							
\end{center}																																							
\end{figure}

\begin{figure}[ht]																																							
\begin{center}																																							
\caption[Ternary vapor-liquid equilibrium phase diagram of the mixture ${\rm CH_4}$ + ${\rm N_2}$ + ${\rm CO_2}$ at 293.19 K and 7.15 MPa: {\Large $\bullet$} present simulation data, $+$ experimental data \cite{vleN2CH4CO2}, --- Peng-Robinson equation of state.]{}																									
\bigskip																																							
\epsfig{file=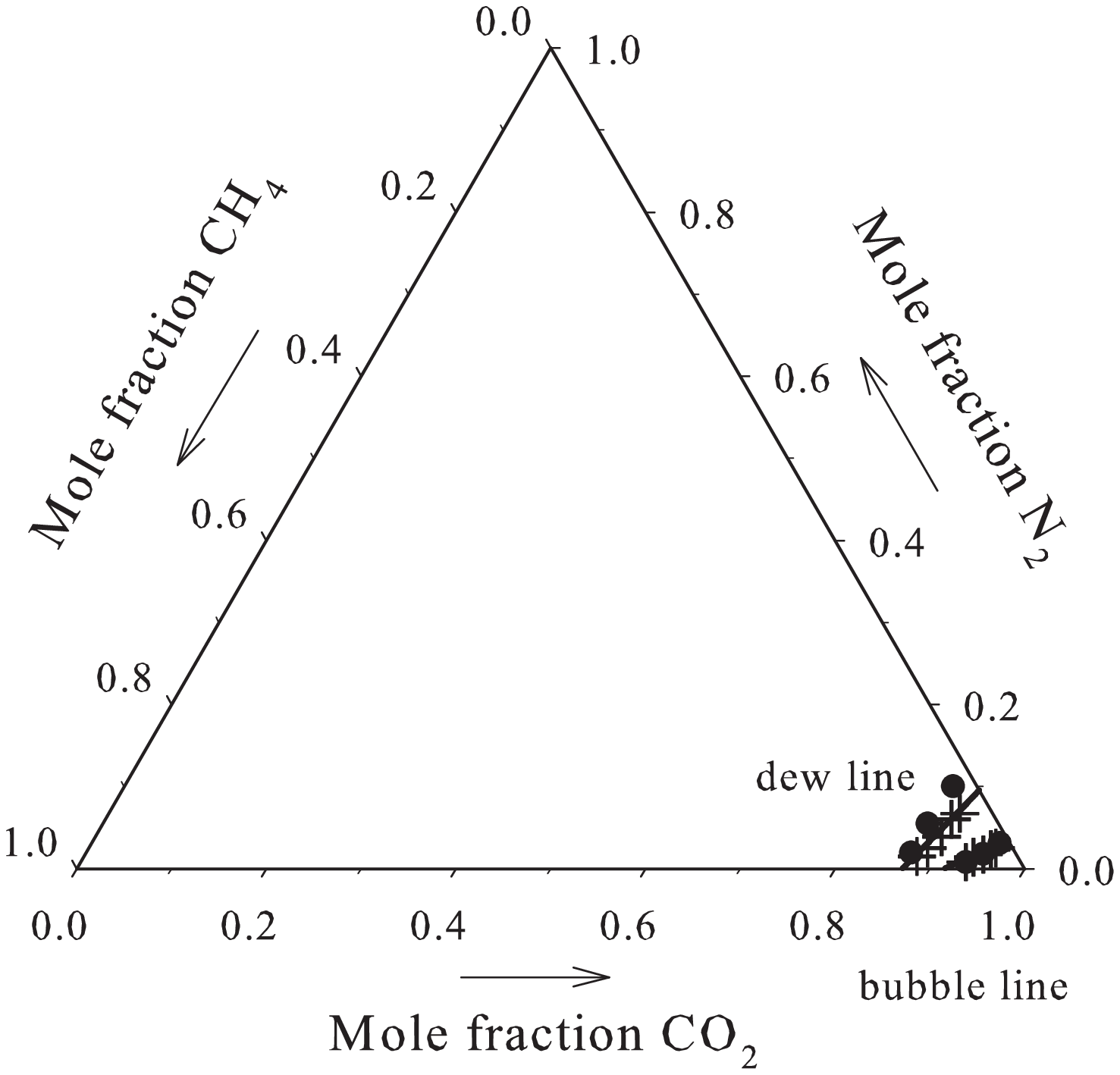,scale=0.56}																																							
\label{ch4_n2_co2} 	 																																							
\end{center}																																							
\end{figure}

\begin{figure}[ht]																																							
\begin{center}																																							
\caption[Ternary vapor-liquid equilibrium phase diagram of the mixture ${\rm CH_4}$ + ${\rm N_2}$ + ${\rm C_2H_6}$ at 220 K and 8 MPa: {\Large $\bullet$} present simulation data, $+$ experimental data \cite {vleN2CH4C2H6}, --- Peng-Robinson equation of state.]{}																																							
\bigskip																																							
\epsfig{file=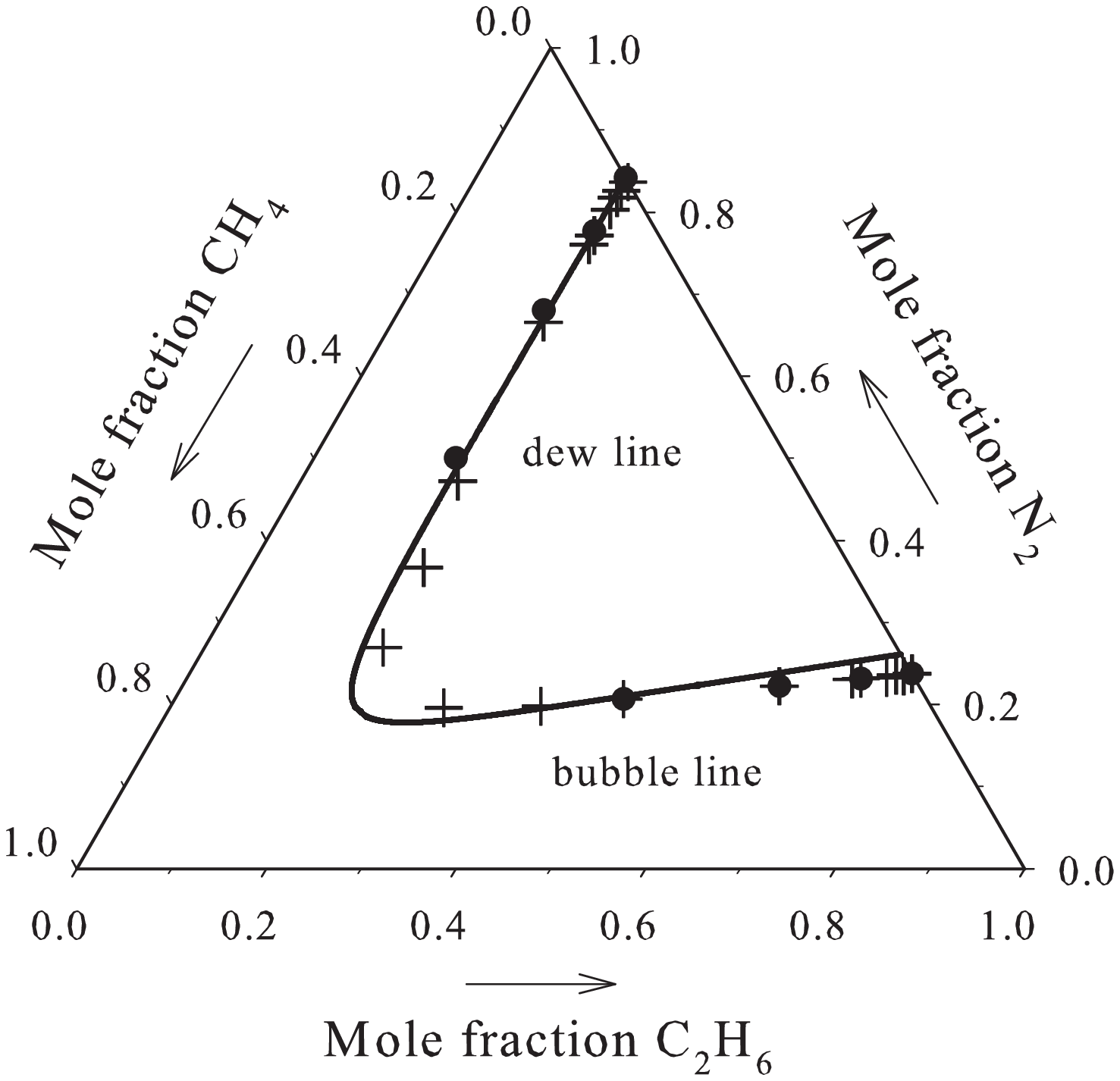,scale=0.56}																																							
\label{ch4_n2_c2h6} 	 																																							
\end{center}																																							
\end{figure}

\begin{figure}[ht]																																							
\begin{center}																																							
\caption[Ternary vapor-liquid equilibrium phase diagram of the mixture ${\rm CH_4}$ + CO + ${\rm CO_2}$ at 223.15 K and 6.7 MPa: {\Large $\bullet$} present simulation data, $+$ experimental data \cite {vleCO2CH4CO}, --- Peng-Robinson equation of state.]{}																																							
\bigskip																																							
\epsfig{file=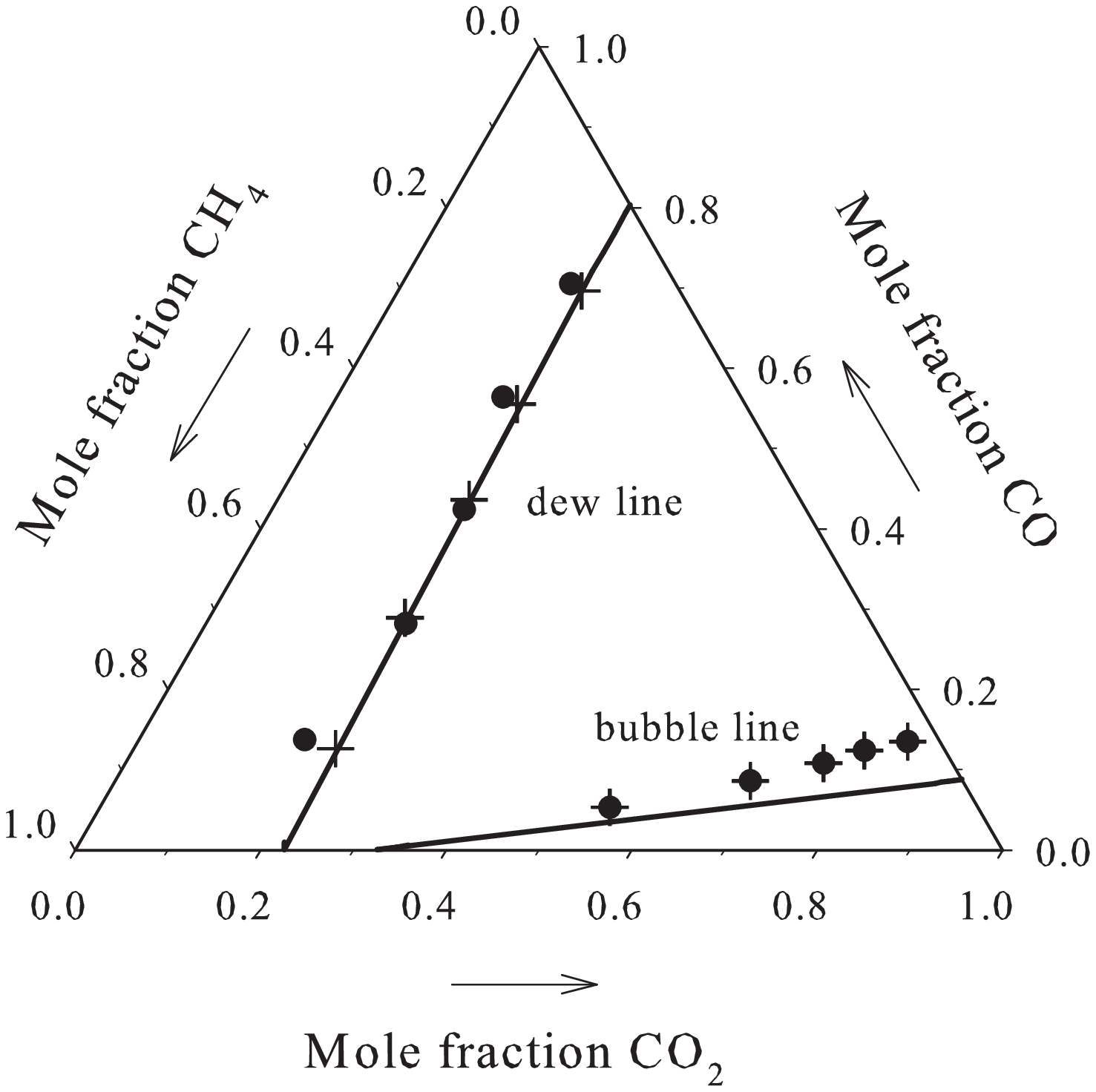,scale=0.56}																																							
\label{ch4_co_co2} 	 																																							
\end{center}																																							
\end{figure}

\begin{figure}[ht]																																							
\begin{center}																																							
\caption[Ternary vapor-liquid equilibrium phase diagram of the mixture ${\rm CH_4}$ + ${\rm CO_2}$ + ${\rm C_2H_6}$ at 230 K and 4.65 MPa: {\Large $\bullet$} present simulation data, $+$ experimental data \cite {vleCH4C2H6CO2}, --- Peng-Robinson equation of state.]{}																																							
\bigskip																																							
\epsfig{file=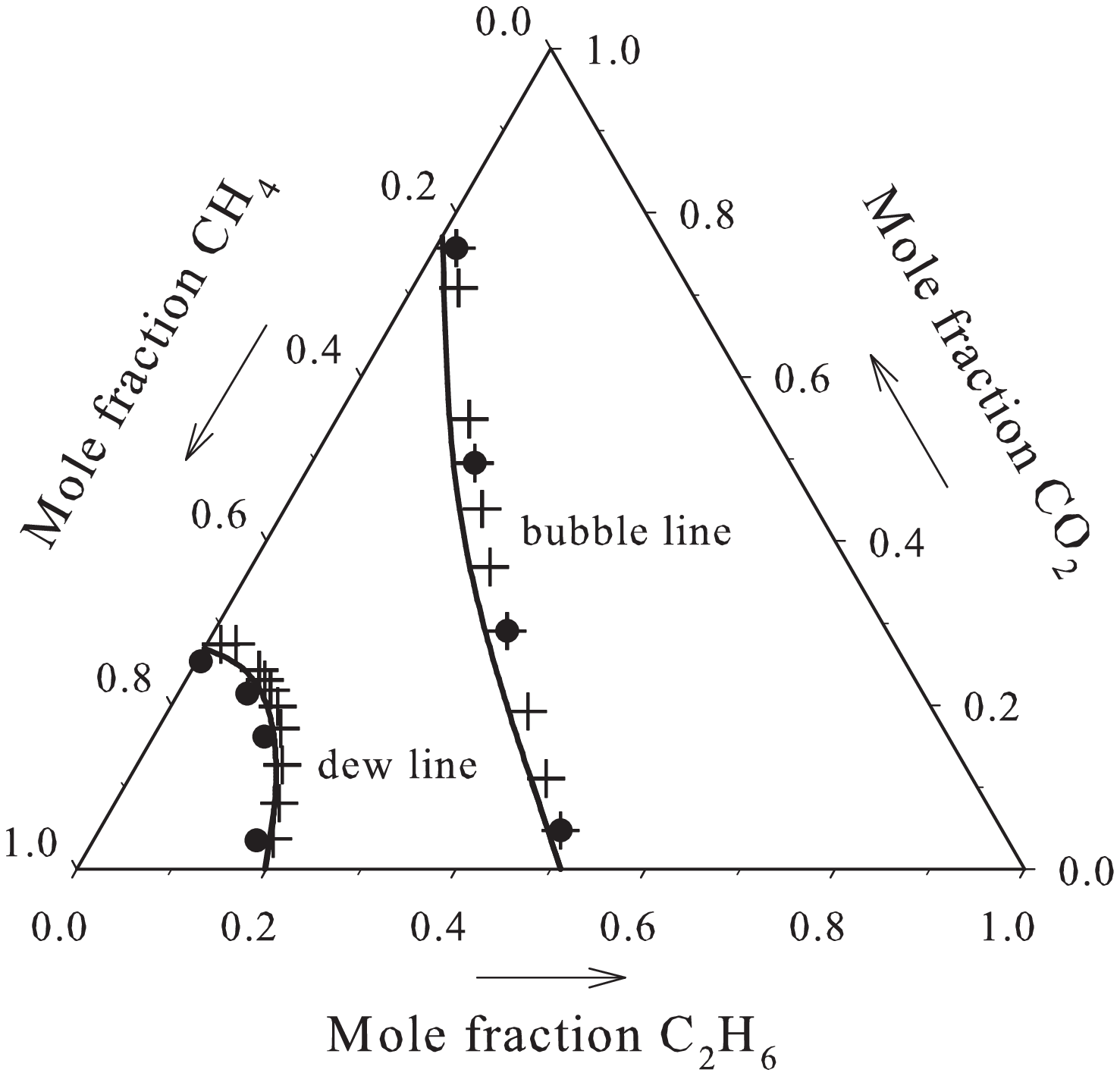,scale=0.56}																																							
\label{ch4_co2_c2h6} 	 																																							
\end{center}																																							
\end{figure}

\begin{figure}[ht]																																							
\begin{center}																																							
\caption[Ternary vapor-liquid equilibrium phase diagram of the mixture ${\rm CH_4}$ + ${\rm C_2H_4}$ + ${\rm C_2H_6}$ at 159.21 K and 0.263 MPa: {\Large $\bullet$} present simulation data, $+$ experimental data \cite {vleCH4C2H4C2H6}, --- Peng-Robinson equation of state.]{}																																							
\bigskip																																							
\epsfig{file=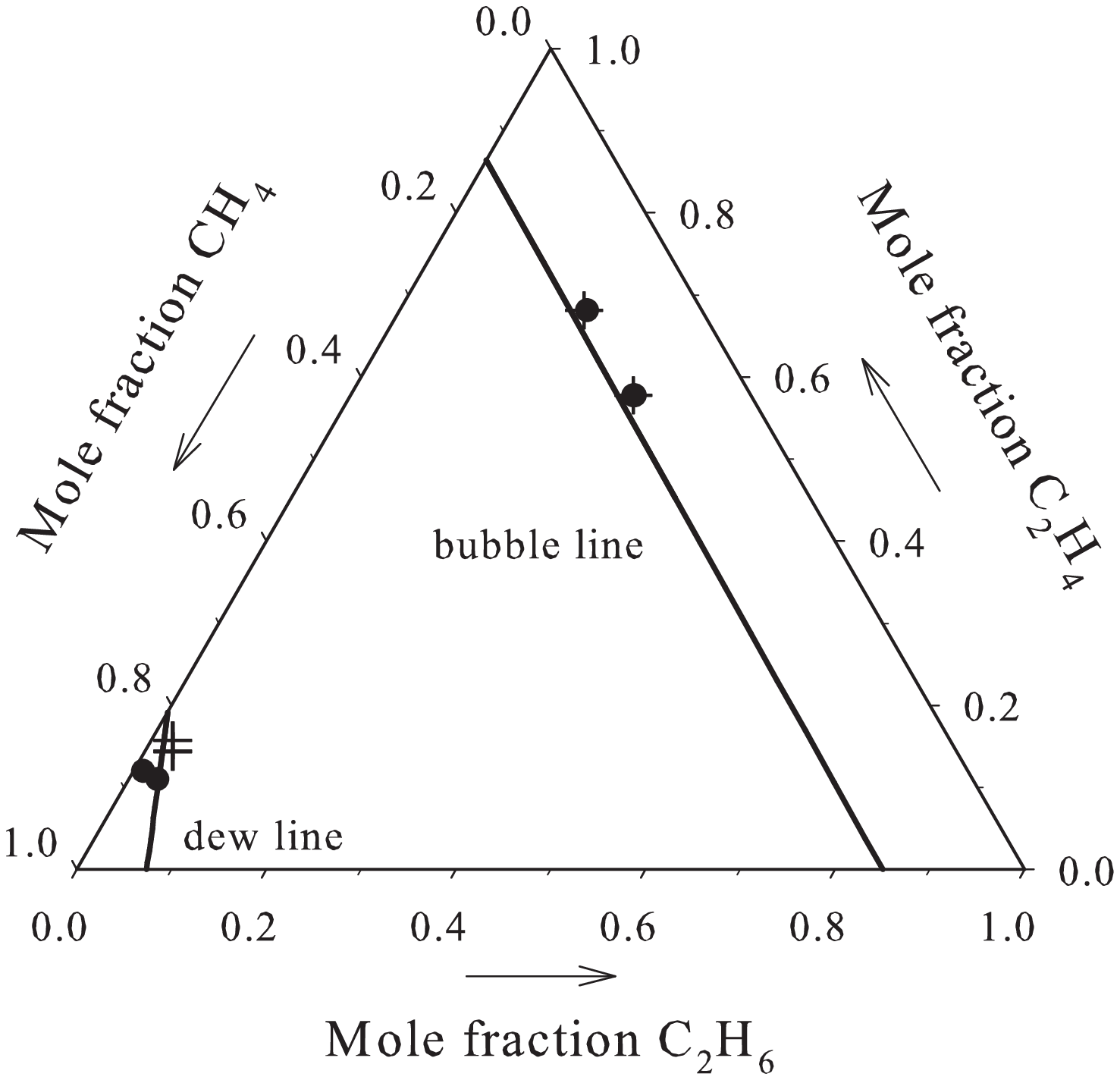,scale=0.56}																																							
\label{ch4_c2h4_c2h6} 	 																																							
\end{center}																																							
\end{figure}

\begin{figure}[ht]																																							
\begin{center}																																							
\caption[Ternary vapor-liquid equilibrium phase diagram of the mixture ${\rm N_2}$ + ${\rm O_2}$ + ${\rm CO_2}$ at 232.85 K and 12.4 MPa: {\Large $\bullet$} present simulation data, $+$ experimental data \cite {vleCO2O2N2}, --- Peng-Robinson equation of state.]{}																																							
\bigskip																																							
\epsfig{file=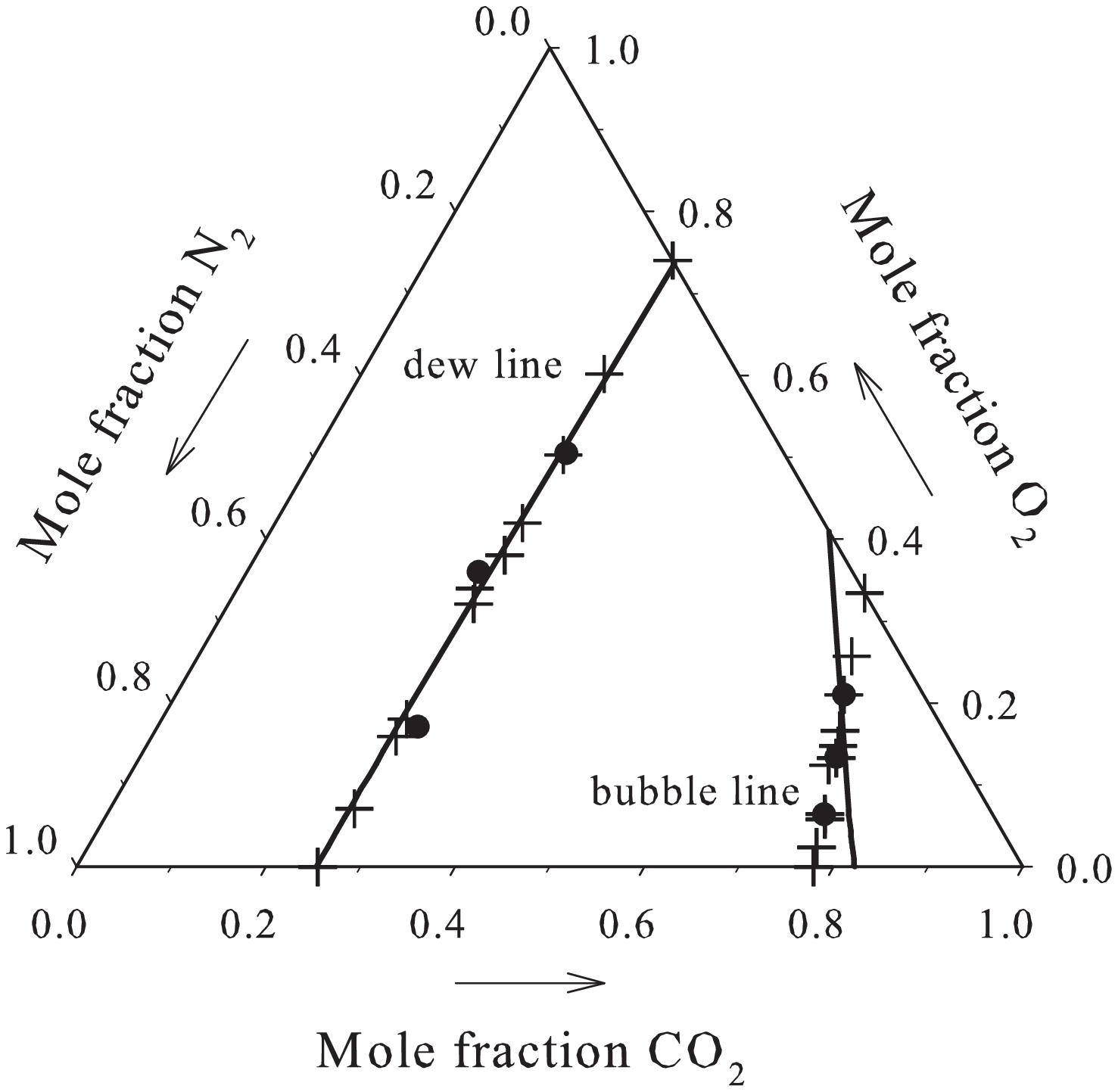,scale=0.56}																																							
\label{n2_o2_co2} 	 																																							
\end{center}																																							
\end{figure}

\begin{figure}[ht]																																							
\begin{center}																																							
\caption[Ternary vapor-liquid equilibrium phase diagram of the mixture ${\rm N_2}$ + ${\rm CO_2}$ + ${\rm C_2H_6}$ at 220 K and 4 MPa: {\Large $\bullet$} present simulation data, $+$ experimental data \cite{vleCO2C2H6N2}, --- Peng-Robinson equation of state.]{}																																							
\bigskip																																							
\epsfig{file=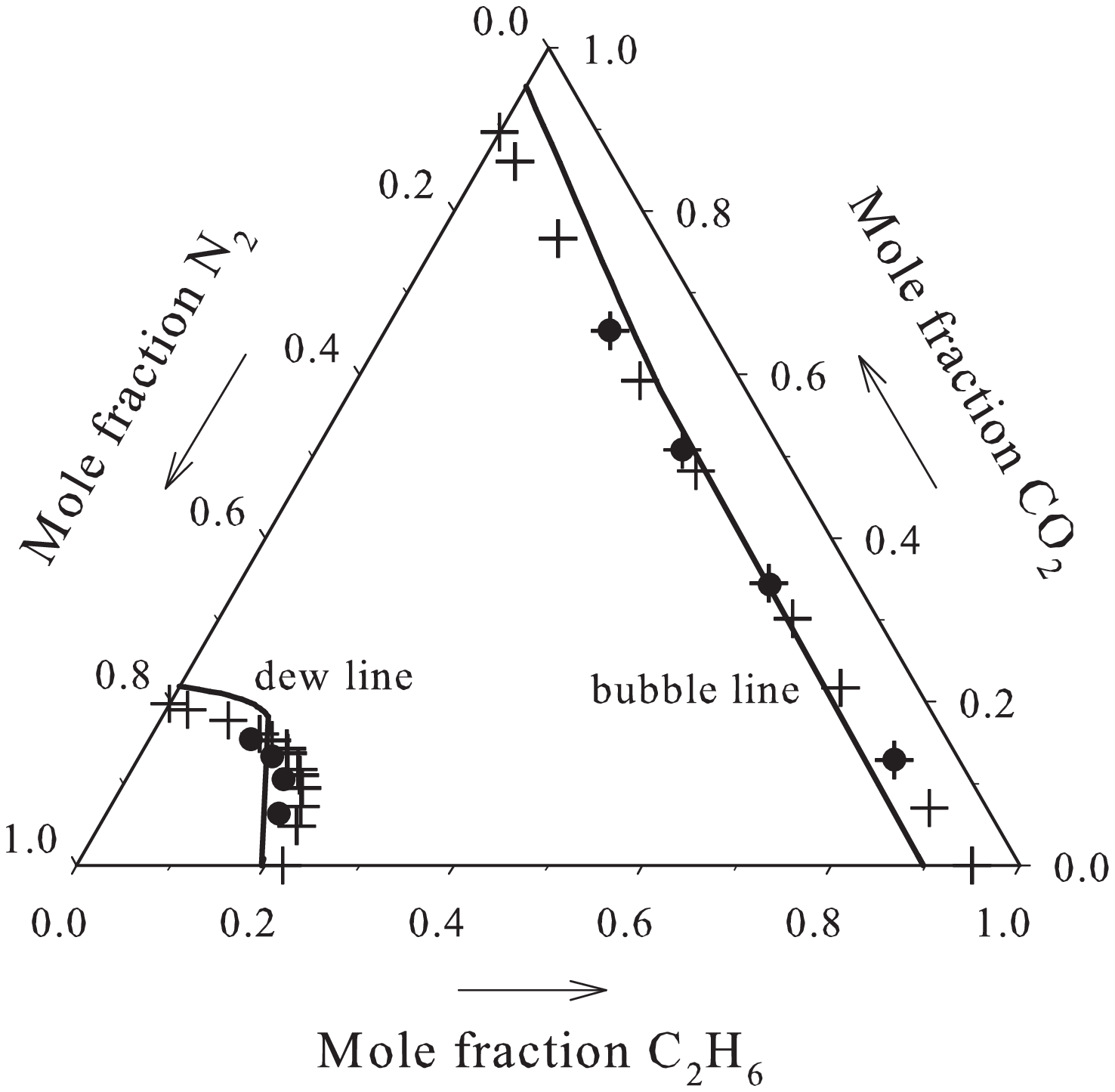,scale=0.56}																																							
\label{n2_co2_c2h6} 	 																																							
\end{center}																																							
\end{figure}

\begin{figure}[ht]																																							
\begin{center}																																							
\caption[Ternary vapor-liquid equilibrium phase diagram of the mixture ${\rm N_2}$ + ${\rm CO_2}$ + R12 at 273.15 K and 5 MPa: {\Large $\bullet$} present simulation data, $+$ experimental data \cite{vleN2CO2R12}, --- Peng-Robinson equation of state.]{}																																							
\bigskip																																							
\epsfig{file=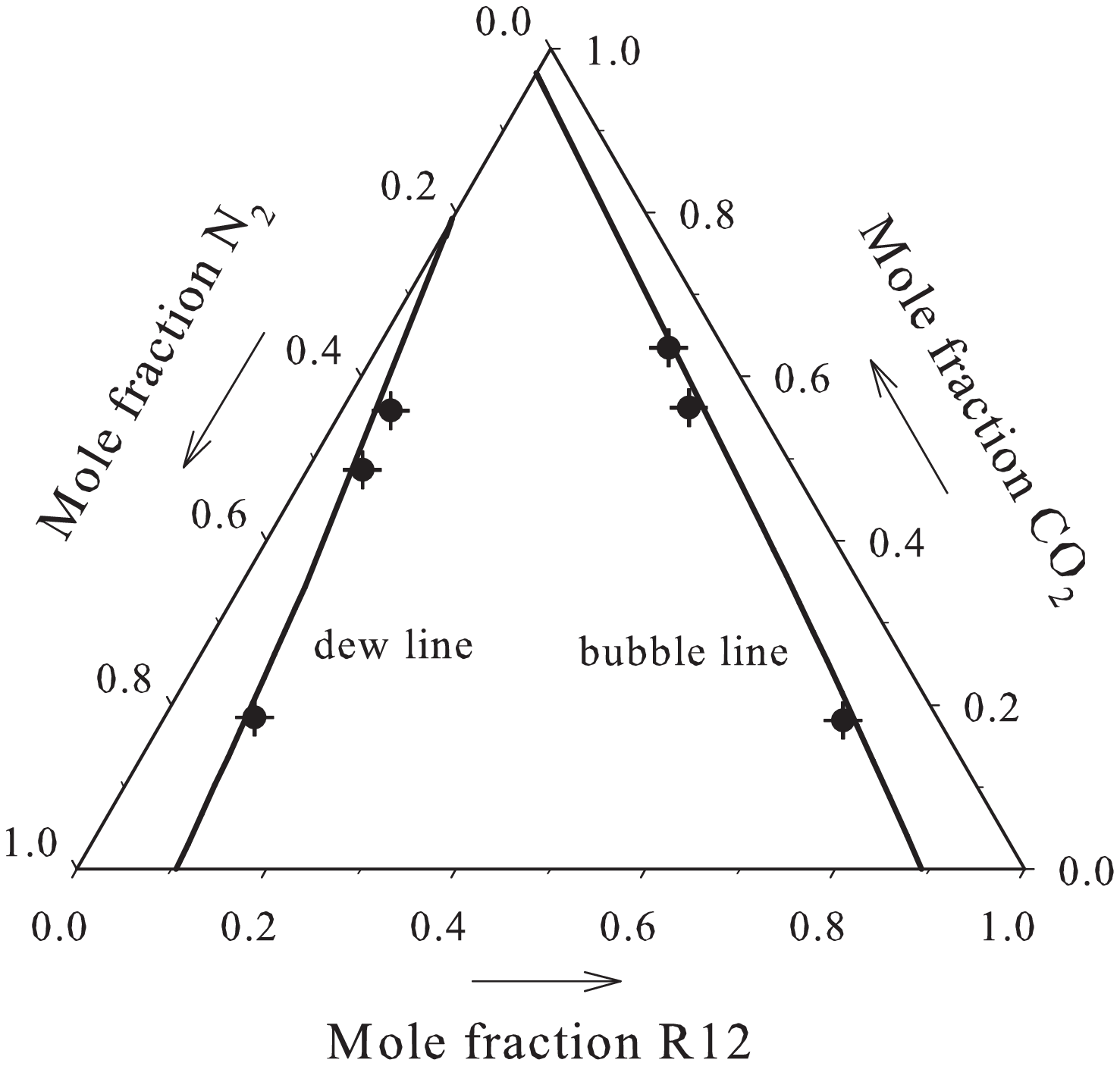,scale=0.56}																																							
\label{n2_co2_r12} 	 																																							
\end{center}																																							
\end{figure}

\begin{figure}[ht]																																							
\begin{center}																																							
\caption[Ternary vapor-liquid equilibrium phase diagram of the mixture ${\rm N_2}$ + ${\rm CO_2}$ + R22 at 273.2 K and 3.083 MPa: {\Large $\bullet$} present simulation data, $+$ experimental data \cite{vleN2CO2R22}, --- Peng-Robinson equation of state.]{}																																							
\bigskip																																							
\epsfig{file=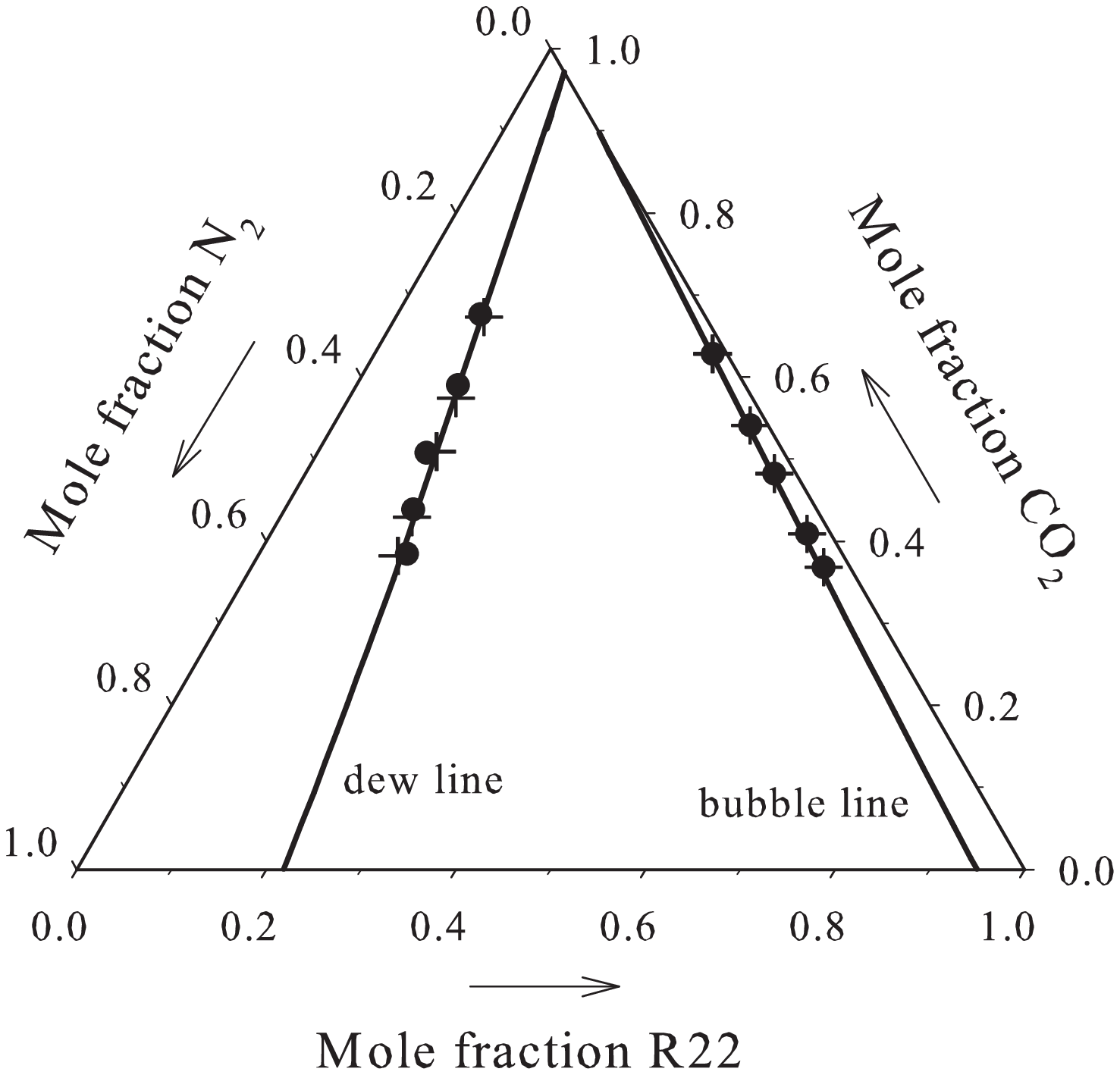,scale=0.56}																																							
\label{n2_co2_r22} 	 																																							
\end{center}																																							
\end{figure}

\begin{figure}[ht]																																							
\begin{center}																																							
\caption[Ternary vapor-liquid equilibrium phase diagram of the mixture ${\rm CO_2}$ + R22 + R142b at 348.8 K and 4.053 MPa: {\Large $\bullet$} present simulation data, $+$ experimental data \cite{vleCO2R22R142b}, --- Peng-Robinson equation of state.]{}																																							
\bigskip																																							
\epsfig{file=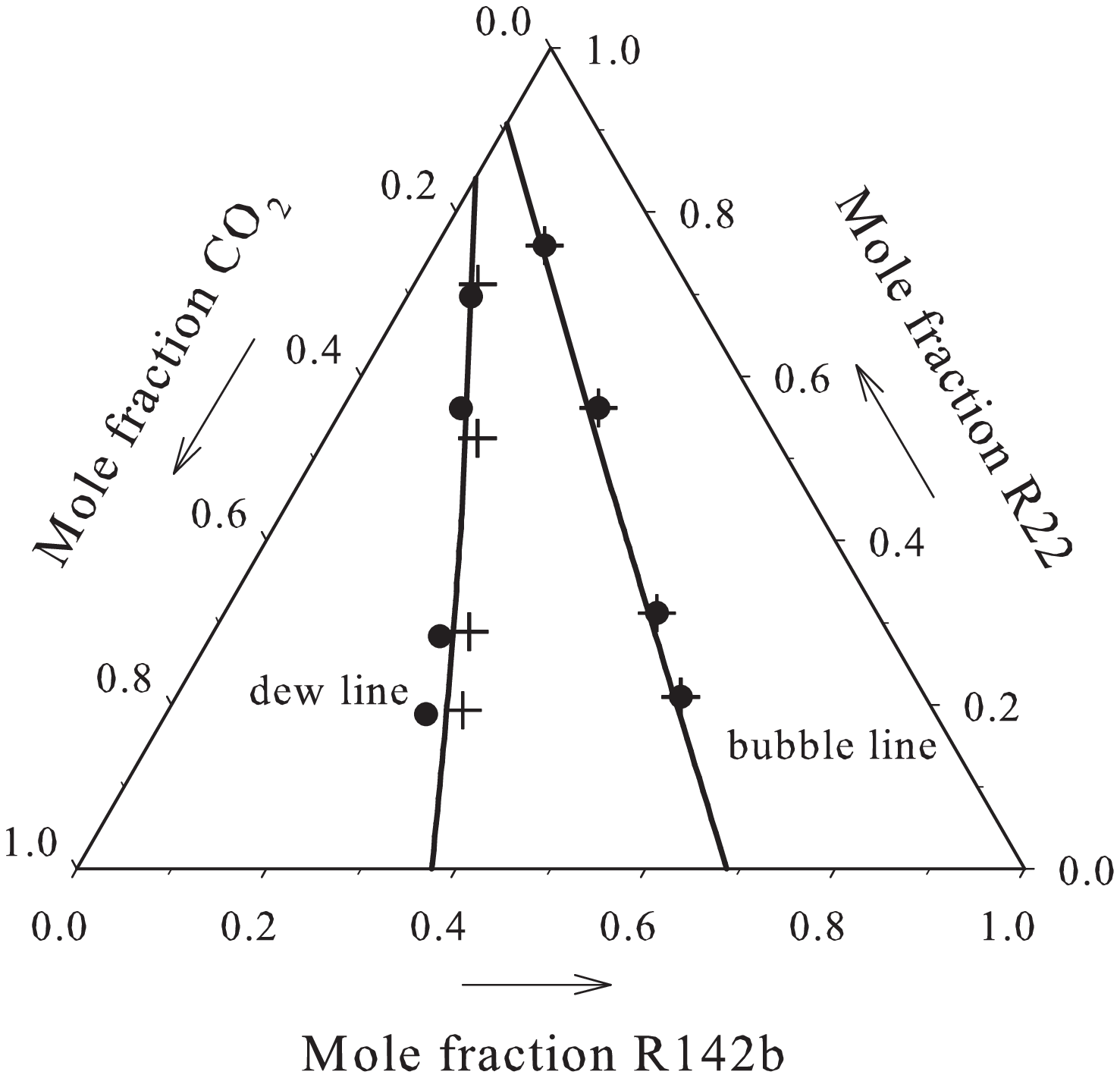,scale=0.56}																																							
\label{co2_r22_r142b} 	 																																							
\end{center}																																							
\end{figure}

\begin{figure}[ht]																																							
\begin{center}																																							
\caption[Ternary vapor-liquid equilibrium phase diagram of the mixture ${\rm CO_2}$ + R142b + R152a at 347.5 K and 4.053 MPa: {\Large $\bullet$} present simulation data, $+$ experimental data \cite{vleCO2R22R142b}, --- Peng-Robinson equation of state.]{}																																							
\bigskip																																							
\epsfig{file=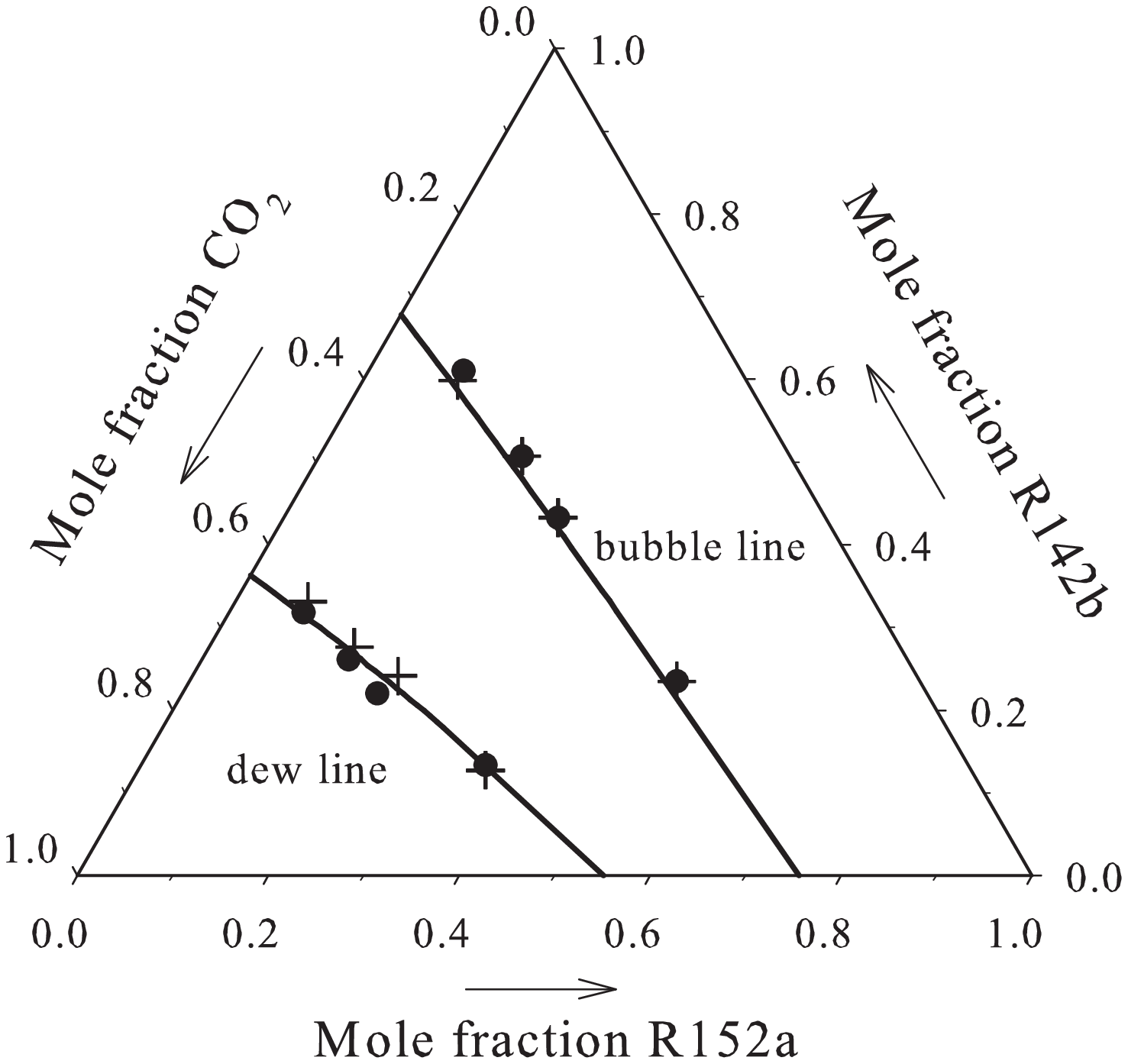,scale=0.56}																																							
\label{co2_r142b_r152a} 																																							
\end{center}																																							
\end{figure}

\begin{figure}[ht]																																							
\begin{center}																																							
\caption[Ternary vapor-liquid equilibrium phase diagram of the mixture ${\rm C_2H_2}$ + ${\rm C_2H_4}$ + ${\rm C_2H_6}$ at 277.6 K and 3.55 MPa: {\Large $\bullet$} present simulation data, $+$ experimental data \cite{vleC2H4C2H6C2H2}, --- Peng-Robinson equation of state.]{}																																							
\bigskip																																							
\epsfig{file=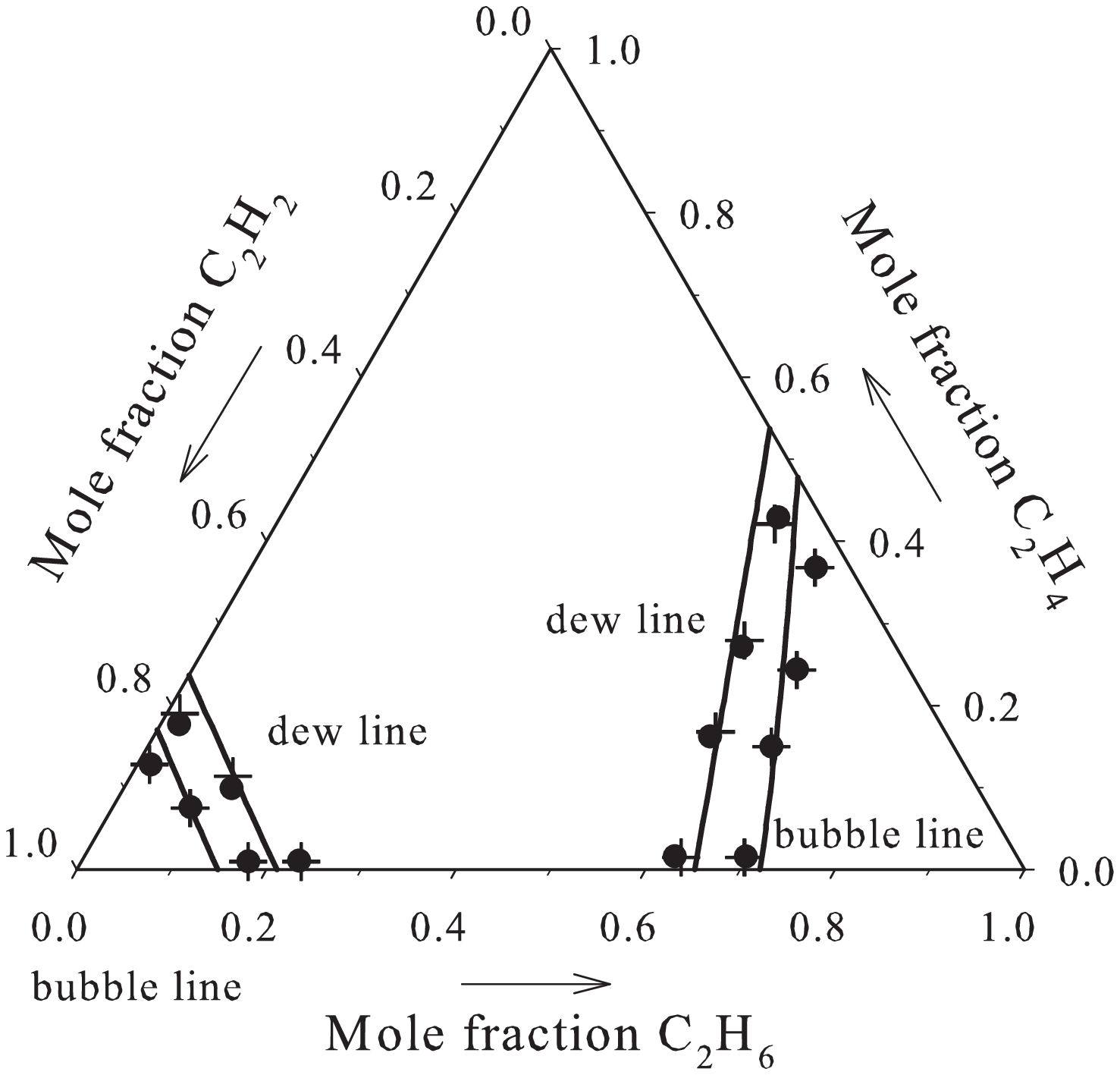,scale=0.56}																																							
\label{c2h2_c2h4_c2h6}  																																							
\end{center}																																							
\end{figure}

\begin{figure}[ht]																																							
\begin{center}																																							
\caption[Ternary vapor-liquid equilibrium phase diagram of the mixture R10 + R20 + R30 at 318.15 K and 0.07 MPa: {\Large $\bullet$} present simulation data, $+$ experimental data \cite{vleCH2Cl2CHCl3CCl4}, --- Peng-Robinson equation of state.]{}																																							
\bigskip																																							
\epsfig{file=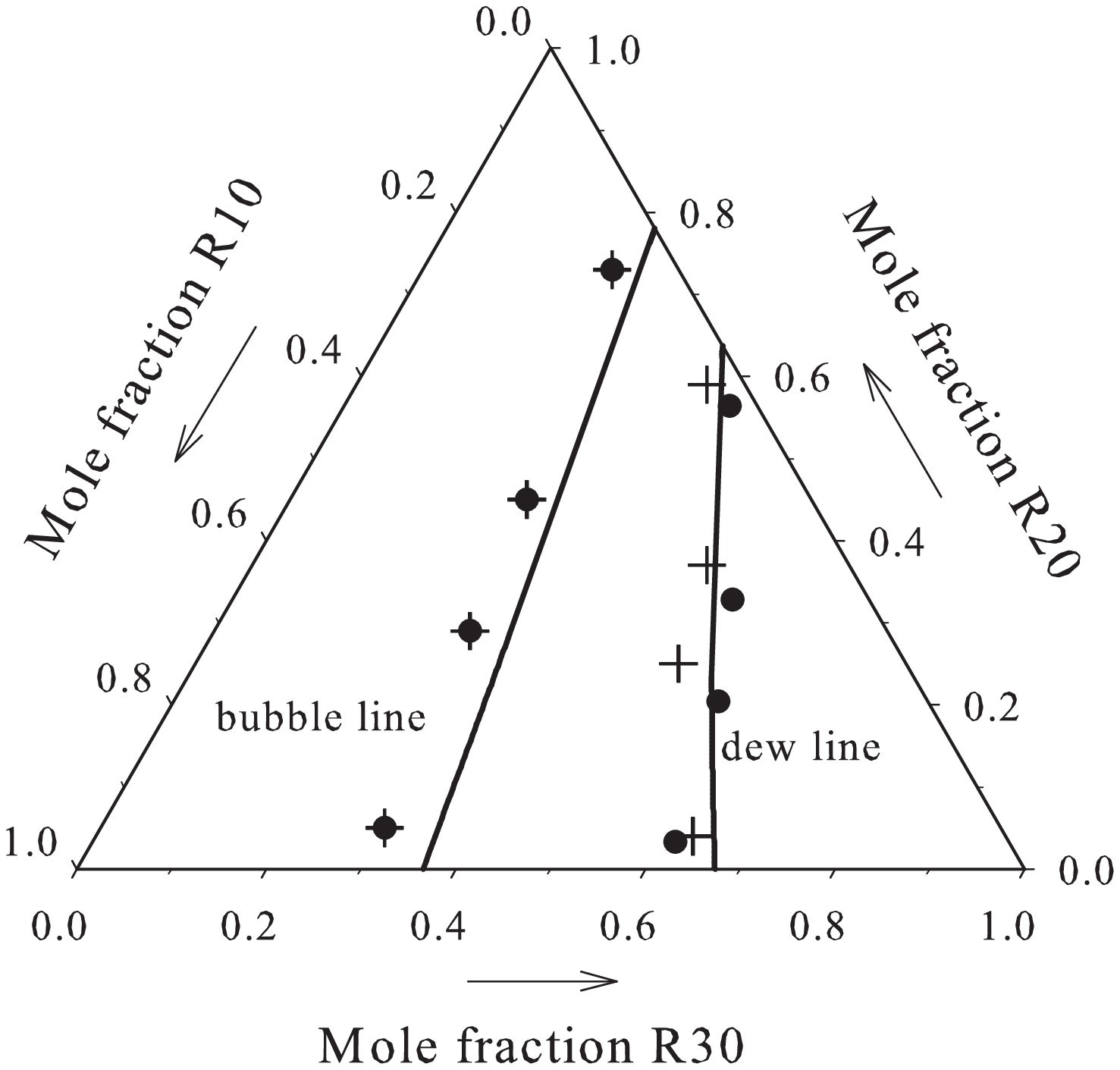,scale=0.56}																																							
\label{r10_r20_r30} 	 																																							
\end{center}																																							
\end{figure}

\begin{figure}[ht]																																							
\begin{center}																																							
\caption[Ternary vapor-liquid equilibrium phase diagram of the mixture R10 + R1110 + R1120 at 358.5 K and 0.101 MPa: {\Large $\bullet$} present simulation data, $+$ experimental data \cite{vleCCl4C2HCl3C2Cl4}, --- Peng-Robinson equation of state.]{}																																							
\bigskip																																							
\epsfig{file=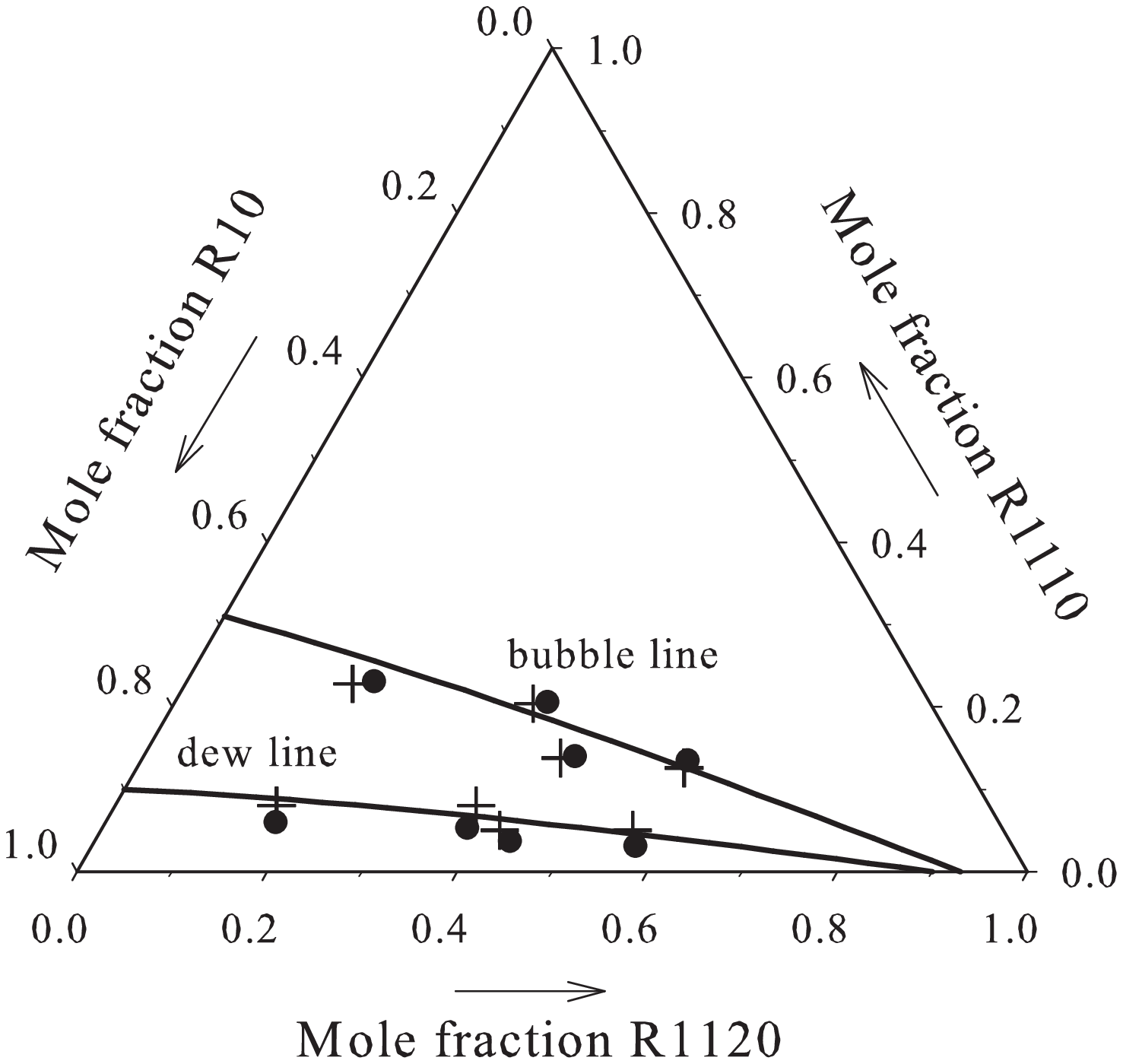,scale=0.56}																																							
\label{r10_r1110_r1120} 																																							
\end{center}																																							
\end{figure}

\begin{figure}[ht]																																							
\begin{center}																																							
\caption[Ternary vapor-liquid equilibrium phase diagram of the mixture R13 + R14 + R23 at 199.8 K and 0.345 MPa: {\Large $\bullet$} present simulation data, $+$ experimental data \cite{vleR14R23R13}, --- Peng-Robinson equation of state.]{}																																							
\bigskip																																							
\epsfig{file=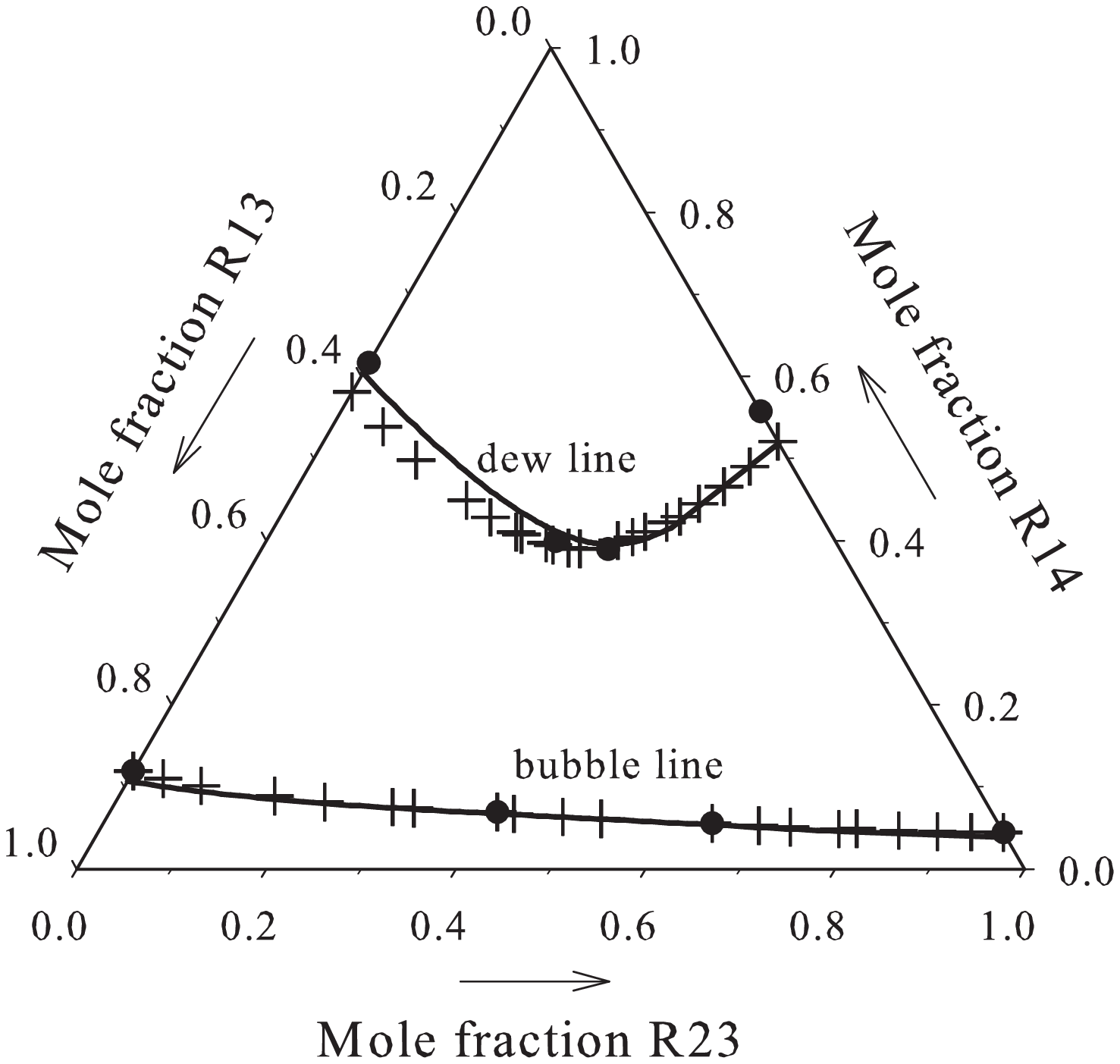,scale=0.56}																																							
\label{r13_r14_r23} 	 																																							
\end{center}																																							
\end{figure}

\begin{figure}[ht]																																							
\begin{center}																																							
\caption[Ternary vapor-liquid equilibrium phase diagram of the mixture R30 + R30B1 + R30B2 at 342.5 K and 0.101 MPa: {\Large $\bullet$} present simulation data, $+$ experimental data \cite{vleCH2Cl2R30B1R30B2}.]{}																																							
\bigskip																																							
\epsfig{file=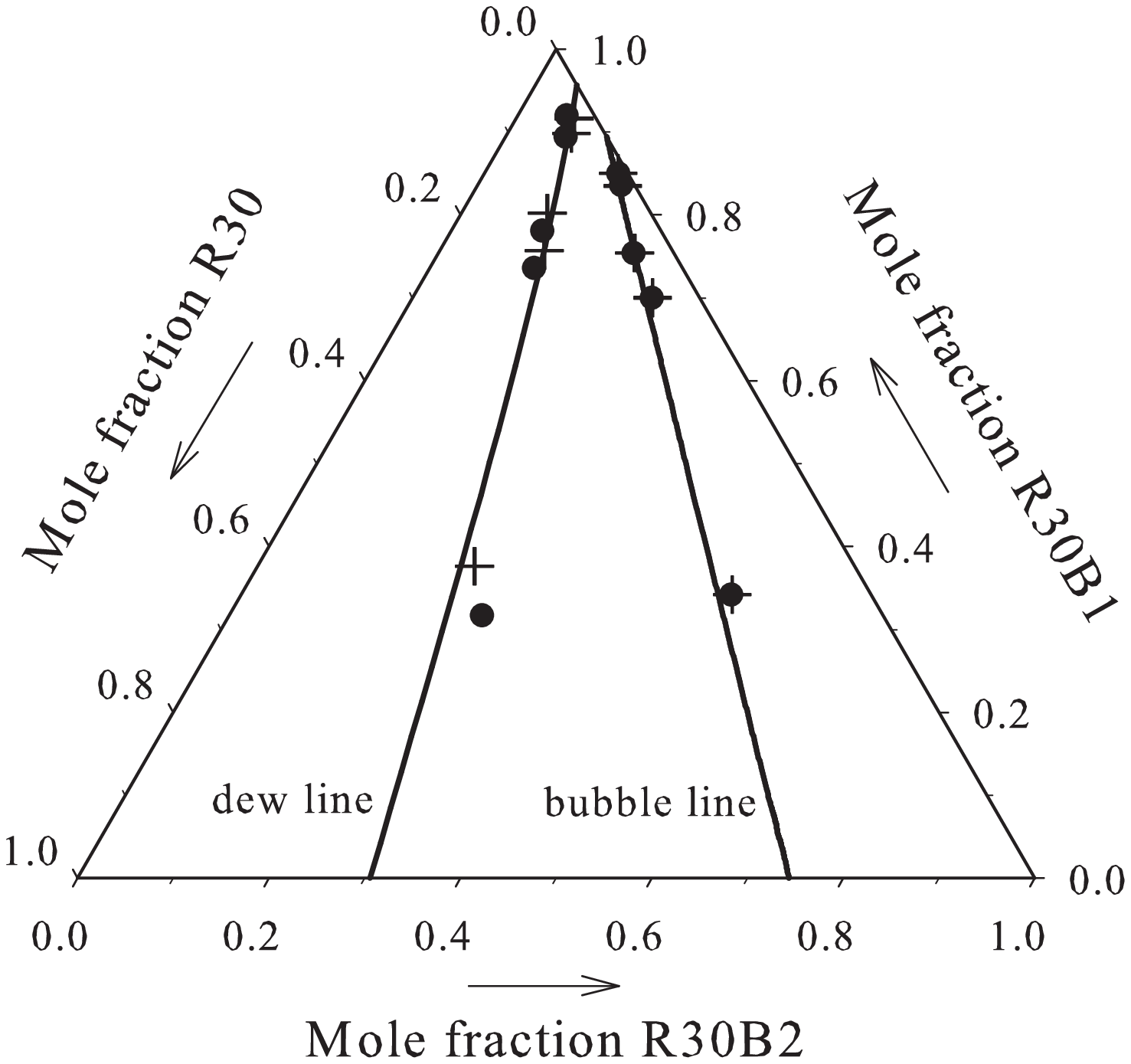,scale=0.56}																																							
\label{r30_r30b1_r30b2} 																																							
\end{center}																																							
\end{figure}

\begin{figure}[ht]																																							
\begin{center}																																							
\caption[Ternary vapor-liquid equilibrium phase diagram of the mixture R32 + R125 + R134a at 333.15 K and 2.43 MPa: {\Large $\bullet$} present simulation data, $+$ experimental data \cite{vleR32R125R134a}, --- Peng-Robinson equation of state.]{}																																							
\bigskip																																							
\epsfig{file=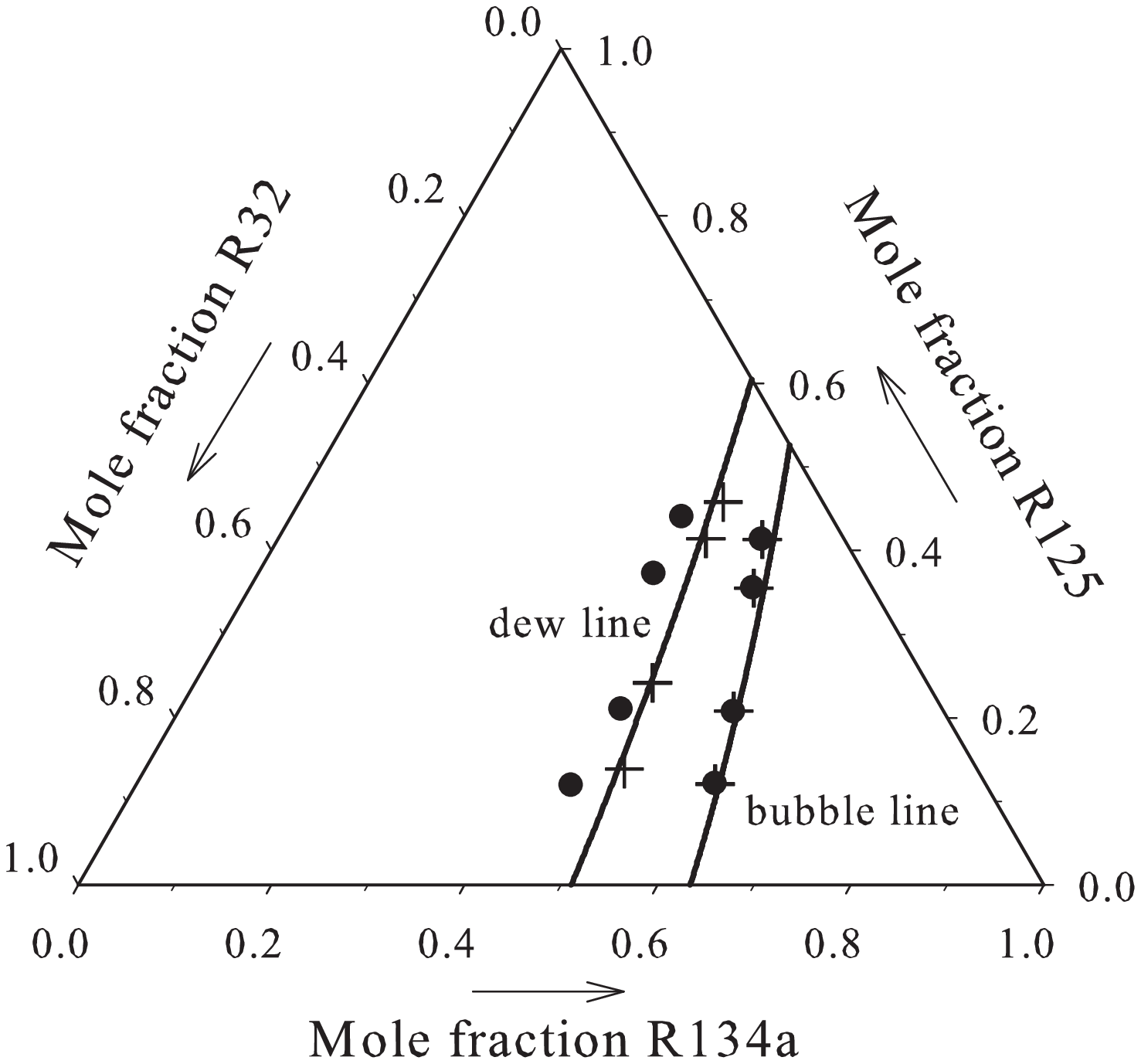,scale=0.56}																																							
\label{r32_r125r_134a}  																																							
\end{center}																																							
\end{figure}

\begin{figure}[ht]																																							
\begin{center}																																							
\caption[Ternary vapor-liquid equilibrium phase diagram of the mixture R140a + R141b + R142b at 323.25 K and 0.25 MPa: {\Large $\bullet$} present simulation data, $+$ experimental data \cite{vleR140aR142bR141b}, --- Peng-Robinson equation of state.]{}
\bigskip																																							
\epsfig{file=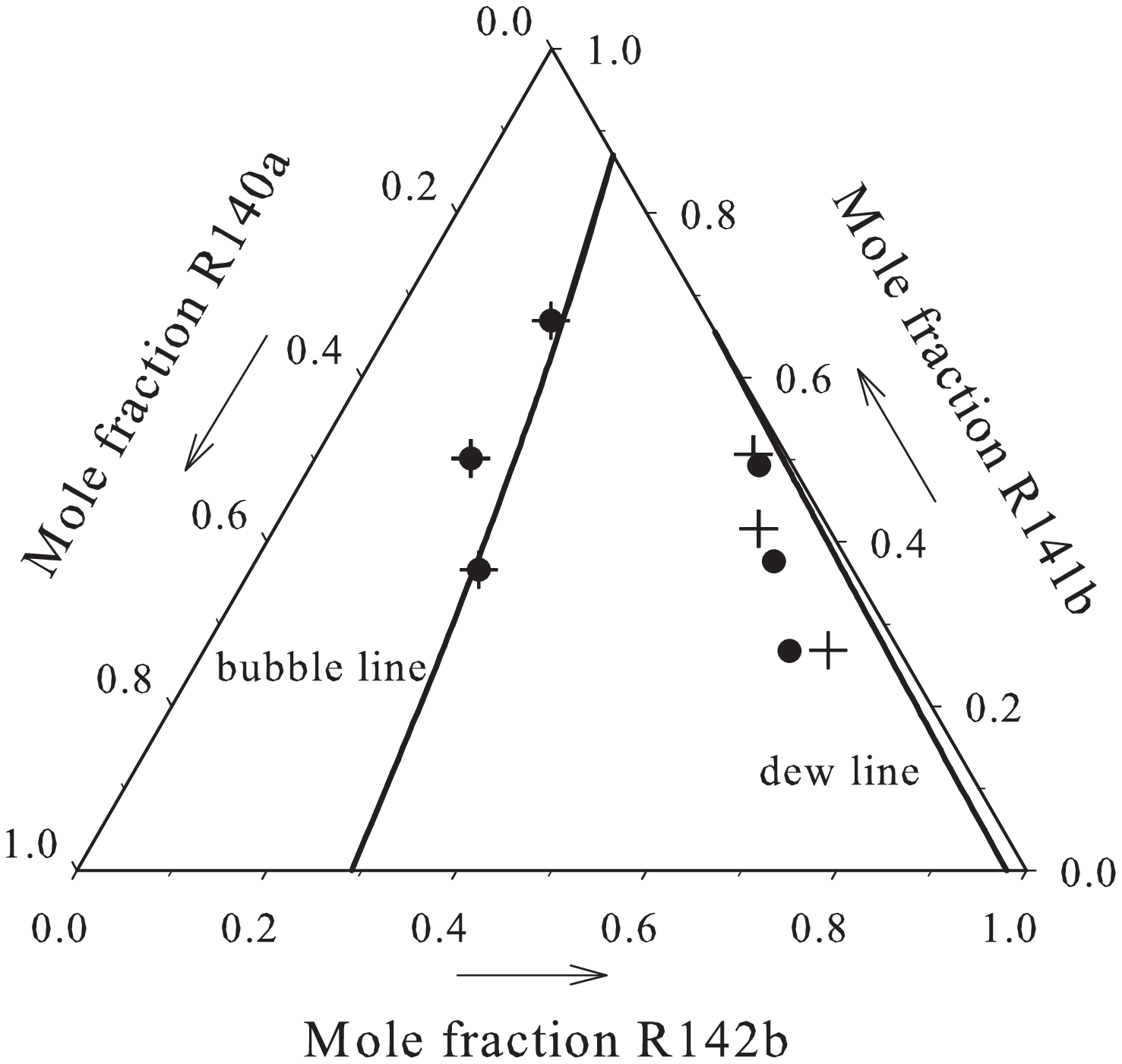,scale=0.56}																																							
\label{r140a_r141b_r142b}																																							
\end{center}																																							
\end{figure}

The results are presented here in ternary plots at constant temperature and pressure, cf. Figures \ref{ar_ch4_n2} to \ref{r140a_r141b_r142b}, covering 23 of the 33 studied ternary mixtures. 
For the remaining ten systems less than two full experimental VLE data points are available for one pair of temperature and pressure so that the results are not presented in figures here, while the numerical comparison to experimental data can be found in the supporting information. 
Full numerical VLE simulation data are given in the supporting information, which also contains saturated densities and heat of vaporization from simulation.

For all predicted VLE properties, an estimate of the statistical uncertainty is provided in the supporting information. 
Due to the fact that the error bars are mostly within symbol size, they were omitted in the figures.

The present assessment was made on the basis of the resulting composition on the saturation lines which can directly be seen in comparison with the experimental data in the phase diagrams of this work.
Note that the simulated vapor pressure in general does not match exactly with the presented pressure, but it is usually very close to it.
The numerical data in the supporting information allows for a direct comparison of the vapor pressure.

Not for all studied systems, the experimental data is sufficient to assess the topology of the saturation lines in the the isobaric-isothermal phase diagrams. 
Most of those, for which this is possible, show a simple topology where one straight bubble line and one straight corresponding dew line connect two binary subsystems,
e.g. Ar + CH$_4$ + N$_2$, cf. Figure \ref{ar_ch4_n2}, N$_2$ + CO$_2$ + R12, cf. Figure \ref{n2_co2_r12}, or CO$_2$ + R142b + R152a, cf. Figure \ref{co2_r142b_r152a}. 
However, three mixtures have markedly curved phase envelopes, i.e. CH$_4$ + CO$_2$ + C$_2$H$_6$, cf. Figures \ref{ch4_co2_c2h6}, N$_2$ + CO$_2$ + C$_2$H$_6$, cf. Figure \ref{n2_co2_c2h6}, and R13 + R14 + R23, cf. Figure \ref{r13_r14_r23}. 
All three have one azeotropic subsystem \cite{Zbinary}, which however, cannot directly be seen from the figures for the ternary systems shown here.
The phase envelope is also curved for the mixture CH$_4$ + N$_2$ + C$_2$H$_6$, cf. Figure \ref{ch4_n2_c2h6}, which has a ternary critical point. 
Finally, Figure \ref{c2h2_c2h4_c2h6} for the mixture C$_2$H$_2$ + C$_2$H$_4$ + C$_2$H$_6$, shows two pairs of straight saturation lines which also result from the azeotropic behavior of the subsystem C$_2$H$_2$ + C$_2$H$_6$ \cite{Zbinary}.

The temperature range covered in the present study is quite large, 
i.e. from 112 K (Ar + CH$_4$ + N$_2$, cf. Figure \ref{ar_ch4_n2}) to 358.5 K (R10 + R1110 + R1120, cf. Figure \ref{r10_r1110_r1120}). 
The same holds for the pressure range, i.e. from 0.07 MPa (R10 + R20 + R30, cf. Figure \ref{r10_r20_r30}) to 12.4 MPa (N$_2$ + O$_2$ + CO$_2$, cf. Figure \ref{n2_o2_co2}).
For most mixtures, experimental data was available only for one pair of temperature and pressure, however, for 11 ternary systems either two (10) or three (1) pairs were simulated.
Thereby, the largest temperature interval was 50 K (N$_2$ + CO$_2$ + C$_2$H$_6$, cf. Figure \ref{n2_co2_c2h6} and supporting information) 
and the largest pressure interval was 7.23 MPa (N$_2$ + O$_2$ + CO$_2$, cf. Figure \ref{n2_o2_co2} and supporting information). 

In general, it can be stated that the agreement between simulation and experiment is very satisfactory. 
Both qualitatively and quantitatively, an excellent match was found in almost all cases.

Only for the mixture N$_2$ + R13 + R14, the VLE could not be simulated at the conditions for which experimental data \cite{vleN2R14R13} is available.
Particularly the temperature of 77.8 K, which is well below the pure substance triple temperature of both R13 (92 K) and R14 (89.5 K), posed a problem during simulation as it is in immediate vicinity to solidification.  
It should be pointed out that the employed molecular models \cite{vrabec0112,vrabec0311} were neither optimized nor evaluated with respect to the triple line. 

Also results from Peng-Robinson EOS with adjusted binary parameter $k_{ij}$ are shown. 
This model is known to be a good correlation tool, making it a workhorse in process engineering, it performs satisfactory in many cases too. 
Within the 23 examples presented here, three cases can be identified where significant deviations to the experimental data can be seen, 
i.e. N$_2$ + O$_2$ + CO$_2$, cf. Figure \ref{n2_o2_co2}, R10 + R20 + R30, cf. Figure \ref{r10_r20_r30} and R140a + R141b + R142b, cf. Figure \ref{r140a_r141b_r142b}.
Finally, for Ar + N$_2$ + O$_2$, cf. Figure \ref{ar_n2_o2}, the Peng-Robinson EOS deviates substantially from the remaining two data sets.

\section{CONCLUSION}
It was shown that molecular modeling and simulation is a reliable and robust approach to obtaining VLE of ternary mixtures. 
To verify this, a total of 33 ternary mixtures were studied by molecular simulation with the Grand Equilibrium method. 
This method was found to be well suited for simulations of ternary VLE. 

The pure component models used in the present study were adjusted to pure component VLE data in previous work. 
For the binary mixtures, one state-independent parameter was adjusted to binary VLE data in other previous work as well.
Due to the fact that pairwise additive potentials were employed and no adjustment of binary parameters to ternary data was carried out, all results of the present study on ternary systems are predictive.
An excellent agreement between the predictions and the experimental data was observed in most cases.
Reliable predictions can also be expected for VLE of mixtures containing more than three components.

Compared to the Peng-Robinson EOS, molecular modeling and simulation is found to yield superior predictions. 

Due to their numerical efficiency and accuracy, the molecular mixture models studied here are also well suited for simulations on a larger scale to investigate processes like evaporation, adsorption, flow etc.

\bigskip
\bigskip
\noindent {\bf ACKNOWLEDGEMENTS} 

We gratefully acknowledge Deutsche Forschungsgemeinschaft for funding this project. 
The simulations are performed on the national super computer NEC SX-8 at the High Performance Computing Center Stuttgart (HLRS) and on the HP X6000 super computer at the Steinbuch Centre for Computing, Karlsruhe.
Furthermore, we would like to thank Xijun Fu, Tania Granados and Ping Lu for performing numerous simulation runs. 

\bigskip
\bigskip
\noindent {\bf SUPPORTING INFORMATION}

The supporting information contains the full numerical simulation data in comparison to the experiment for all studied ternary mixtures. 
Furthermore, the data is presented in ternary vapor-liquid phase diagrams. Note that for five systems figures are omitted, due to the fact that they either could not be simulated (N$_2$ + R13 + R14) or only a single bubble point is available from experiment (R12 + R113 + R152a, R22 + R124 + R142b, R23 + R113 + R114 and R125 + R134a + R161).

This material is available free of charge via the Internet at http://pubs.acs.org. 

\clearpage
{\bf APPENDIX, SIMULATION DETAILS}

\bigskip

The technical simulation details of the present calculations are similar to those published in \cite{Zbinary,vrabec0321}. 
A center-center cut-off radius of 17.5 \r{A} was used for the explicit evaluation of the intermolecular interactions.
The Lennard-Jones tail corrections for internal energy, pressure, and chemical potential were calculated employing angle averaging as proposed by Lustig \cite{lustig}. 
Long-range corrections for the dipolar part of the potential models were made with the reaction field method \cite{barker7378,saager9127}. 
The quadrupolar interaction needs no long range correction as it disappears by orientational averaging. 
The same holds for the mixed polar interaction between dipoles and quadrupoles, cf. Weingerl  et al. \cite{weingerl202}.

VLE were obtained with the Grand Equilibrium method \cite{vrabec15}.
Depending on thermodynamic conditions, two levels of computational effort were employed: 

{\it (A)} In simple cases (e.g. CH$_4$ + CO$_2$ + C$_2$H$_6$, CO$_2$ + R142b + R152a and R13 + R14 + R23) VLE can be obtained with small statistical uncertainties sampling $N=864$ molecules for the liquid phase and about $500$ molecules for the vapor phase. 
Liquid simulation runs were carried out using molecular dynamics with 200~000 time steps, vapor simulation runs were performed using the Monte Carlo technique with 200~000 cycles. 
Within one cycle, $N$ attempts to translate or rotate, and two attempts to insert or delete molecules were sampled. 
The chemical potentials were calculated by Widom's insertion technique \cite{widom1} using 3456 test molecules each time step. 

{\it (B)} In difficult cases (e.g. Ar + N$_2$ + O$_2$, R10 + R20 + R30 and R30 + R30B1 + R30B2), where experimental data is present only for highly dense 
strongly polar liquid phases and the vapor pressure is usually low, the more elaborate gradual insertion scheme had to be employed to obtain the chemical potentials
in the liquid.  

The gradual insertion method is an expanded ensemble method \cite{shevkunov8824} based on the Monte Carlo technique. 
The version as proposed by Nezbeda and Kolafa \cite{nezbeda9139}, extended to the $N\!pT$ ensemble \cite{vrabec0243}, was used in case {\it (B)}. 
In comparison to Widom's insertion technique, where full molecules are inserted into the fluid, gradual insertion introduces one fluctuating molecule that undergoes changes in a predefined set of discrete states of coupling with all other molecules constituting the fluid. Preferential sampling is done in the vicinity of the fluctuating molecule. 
This concept leads to considerably improved accuracy of the residual chemical potential. 
Gradual insertion simulations were performed with $N=864$ molecules in the liquid phase. 
Starting from a face-centered cubic lattice arrangement, every simulation run was given $5 000$ Monte Carlo cycles to equilibrate. 
Data production was performed over 100 000 Monte Carlo cycles. 
One Monte Carlo cycle is defined here as $N$ trial translations, $\left(2/3\right)N$ trial rotations, and one trial volume change. Further simulation parameters for runs with the gradual insertion method were taken from Vrabec et al. \cite{vrabec0243}.

\clearpage


\noindent {\bf LIST OF SYMBOLS} \\[-1cm]

\subsection*{Latin Letters}
\begin {tabbing}
\hspace{2.5cm}    \= \hspace{13cm}  \kill
   $ k_{ij}  $ \> binary parameter of the Peng-Robinson equation of state	\\[-0.12cm]
   $ L       $ \> elongation 							\\[-0.12cm]
   $ p       $ \> pressure							\\[-0.12cm]
   $ Q       $ \> quadrupolar momentum						\\[-0.12cm]
   $ T       $ \> temperature							\\[-0.12cm]
   $ x       $ \> mole fraction in liquid phase					\\[-0.12cm]
   $ y       $ \> mole fraction in vapor phase					\\
\end{tabbing}

\subsection*{Greek Letters}
\begin {tabbing}
\hspace{2.5cm}    \= \hspace{13cm} \kill
   $ \epsilon $ \> Lennard-Jones energy parameter 					\\[-0.12cm]
   $ \mu      $ \> dipolar momentum 							\\[-0.12cm]
   $ \xi      $ \> binary interaction parameter 					\\[-0.12cm]
   $ \sigma   $ \> Lennard-Jones size parameter 					\\[-0.12cm]
\end{tabbing}

\subsection*{Subscripts}
\begin {tabbing}
\hspace {2.5cm}  \= \hspace{13cm} \kill
%
     A         \> related to component A		\\[-0.12cm]
     B         \> related to component B		\\[-0.12cm]
     C         \> related to component C		\\[-0.12cm]
   $ i       $ \> related to component $i$		\\[-0.12cm]
   $ ij      $ \> related to components $i$ and $j$	\\[-0.12cm]
   $ j       $ \> related to component $j$		\\[-0.12cm]
\end{tabbing}
\subsection*{Superscripts}
\begin {tabbing}
\hspace {2.5cm}  \= \hspace{13cm} \kill
%
   $ exp     $ \> experimental data 	\\[-0.12cm]
   $ sim     $ \> simulation data 	\\
\end{tabbing}
\subsection*{Abbreviations}
\begin {tabbing}
\hspace {2.5cm}  \= \hspace{13cm} \kill
%
     1CLJ      \> one-center Lennard-Jones 			\\[-0.12cm]
     1CLJD     \> one-center Lennard-Jones plus point dipole 	\\[-0.12cm]
     2CLJ      \> two-center Lennard-Jones 			\\[-0.12cm]
     2CLJD     \> two-center Lennard-Jones plus point dipole 	\\[-0.12cm]
     2CLJQ     \> two-center Lennard-Jones plus point quadrupole \\[-0.12cm]
     DDB       \> Dortmunder Datenbank 				\\[-0.12cm]
     EOS       \> equation of state 				\\[-0.12cm]
     VLE       \> vapor-liquid equilibria 			\\
\end{tabbing}
\clearpage


\addcontentsline{toc}{chapter}{\numberline{}Bibliography}

\clearpage


\begin{table}[ht]
\noindent
\caption[]{\baselineskip25pt
List of the 35 components studied in the present work. The model parameters were taken from \cite{vrabec0112,vrabec0311}.}
\label{pureall}
\bigskip
\begin{center}
\begin{tabular}{lr|lr} \hline\hline

Component & CAS RN  & Component & CAS RN        \\ \hline

{\bf Non-polar, 1CLJ}  		   &		  & R141b ($\rm CH_3\!\!-\!\!CFCl_2$)	&  1717-00-6	\\[-0.2cm]      	     
Ar 	     			   &  13965-95-2  & R142b ($\rm CH_3\!\!-\!\!CF_2Cl$)	&  75-68-3	\\[-0.2cm]          
$\rm CH_4$   			   &  74-82-8	  & R143a ($\rm CH_3\!\!-\!\!CF_3$)	&  420-46-2	\\[-0.2cm]          
{\bf Dipolar, 1CLJD}   		   &		  & R152a ($\rm CH_3\!\!-\!\!CHF_2$)	&  75-37-6	\\[-0.2cm]          
R30 ($\rm CH_2Cl_2$)		   &  75-09-2     & R161 ($\rm CH_2F\!\!-\!\!CH_3$)	&  353-36-3	\\[-0.2cm]          
R30B2 ($\rm CH_2Br_2$)             &  74-95-3     & {\bf Quadrupolar, 2CLJQ}		&		\\[-0.2cm]          
R32 ($\rm CH_2F_2$)                &  75-10-5     & $\rm N_2$				&  7727-37-9	\\[-0.2cm]          
{\bf Dipolar, 2CLJD}		   &		  & $\rm O_2$				&  7782-44-7	\\[-0.2cm]          
CO	                           &  630-08-0	  & $\rm CO_2$  			&  124-38-9	\\[-0.2cm]          
R11 ($\rm CFCl_3$)		   &  75-69-4	  & $\rm C_2H_2$			&  74-86-2	\\[-0.2cm]          
R12 ($\rm CF_2Cl_2$)		   &  75-71-8	  & $\rm C_2H_4$			&  74-85-1	\\[-0.2cm]          
R13 ($\rm CF_3Cl$)		   &  75-72-9 	  & $\rm C_2H_6$			&  74-84-0	\\[-0.2cm]          
R20 ($\rm CHCl_3$)		   &  67-66-3	  & R10 ($\rm CCl_4$)			&  56-23-5	\\[-0.2cm]          
R22 ($\rm CHF_2Cl$)		   &  75-45-6	  & R14 ($\rm CF_4$)			&  75-73-0	\\[-0.2cm]          
R23 ($\rm CHF_3$)		   &  75-46-7	  & R113 ($\rm CFCl_2\!\!-\!\!CF_2Cl$)  &  76-13-1	\\[-0.2cm]          
R30B1 ($\rm CH_2BrCl$)	 	   & 74-97-5	  & R114 ($\rm CF_2Cl\!\!-\!\!CF_2Cl$)  &  76-14-2	\\[-0.2cm]          
R124 ($\rm CHFCl\!\!-\!\!CF_3$)    &  2837-89-0   & R150B2 ($\rm CH_2Br\!\!-\!\!CH_2Br$)&  106-93-4	\\[-0.2cm]          
R125 ($\rm CHF_2\!\!-\!\!CF_3$)    &  354-33-6    & R1110 ($\rm C_2Cl_4$)		&  127-18-4	\\[-0.2cm]      		
R134a ($\rm CH_2F\!\!-\!\!CF_3$)   &  811-97-2    & R1120 ($\rm CHCl\!\!=\!\!CCl_2$)	&  79-01-6	\\[-0.2cm]          
R140a ($\rm CCl_3\!\!-\!\!CH_3$)   &  71-55-6     &	&	 \\ \hline\hline				
   
\end{tabular}
\end{center}
\end{table}
\clearpage

\begin{table}[ht]
\noindent
\caption[]{\baselineskip25pt
List of the 33 studied ternary mixtures and reference to literature on experimental VLE. For systems indicated with \dag, only bubble line data is available from experiment.}
\label{terall}
\bigskip
\begin{center}
\begin{tabular}{llllll} \hline\hline

Ar + ${\rm CH_4}$ + ${\rm N_2}$ 	      	& \cite {vleN2ArCH4} 	& ${\rm N_2}$ + ${\rm CO_2}$ + ${\rm C_2H_6}$	  & \cite{vleCO2C2H6N2}      & R13 + R14 + R23       & \cite{vleR14R23R13}	    \\[-0.01cm]
Ar + ${\rm CH_4}$ + CO      	     	       	& \cite {vleCOArCH4} 	& ${\rm N_2}$ + ${\rm CO_2}$ + R12		  	  & \cite{vleN2CO2R12}       & R22 + R23 + R114  \dag & \cite{vleR23R22R114}    \\[-0.01cm]
Ar + ${\rm CH_4}$ + ${\rm C_2H_6}$	       	& \cite {vleArCH4C2H6}	& ${\rm N_2}$ + ${\rm CO_2}$ + R22		  	  & \cite{vleN2CO2R22}       & R22 + R124 + R142b \dag & \cite{vleR22R124R142b} \\[-0.01cm]
Ar + ${\rm N_2}$ + ${\rm O_2}$      	      & \cite {vleN2ArO2}	& ${\rm N_2}$ + R13 + R14 \dag	     		  & \cite{vleN2R14R13}       & R22 + R142b + R152a   & \cite{vleR22R152aR142b}   \\[-0.01cm]
${\rm CH_4}$ + ${\rm N_2}$ + CO      	      & \cite {vleN2COCH4}	& ${\rm CO_2}$ + R22 + R142b	       		  & \cite{vleCO2R22R142b}    & R23 + R113 + R114 \dag  & \cite{vleR23R114R113}  \\[-0.01cm]
${\rm CH_4}$ + ${\rm N_2}$ + ${\rm CO_2}$       & \cite {vleN2CH4CO2}	& ${\rm CO_2}$ + R142b + R152a         		  & \cite{vleCO2R22R142b}    & R30 + R30B1 + R30B2   & \cite{vleCH2Cl2R30B1R30B2}\\[-0.01cm]
${\rm CH_4}$ + ${\rm N_2}$ + ${\rm C_2H_6}$     & \cite {vleN2CH4C2H6}	& ${\rm C_2H_2}$ + ${\rm C_2H_4}$ +${\rm C_2H_6}$ & \cite{vleC2H4C2H6C2H2}   & R32 + R125 + R134a    & \cite{vleR32R125R134a}   \\[-0.01cm]
${\rm CH_4}$ + CO + ${\rm CO_2}$      	      & \cite {vleCO2CH4CO}	& R10 + R20 + R30		       		        & \cite{vleCH2Cl2CHCl3CCl4}& R32 + R125 + R143a    & \cite{vleR32R125R143a}   \\[-0.01cm]
${\rm CH_4}$ + ${\rm CO_2}$ + ${\rm C_2H_6}$    & \cite {vleCH4C2H6CO2}	& R10 + R1110 + R1120		       		  & \cite{vleCCl4C2HCl3C2Cl4}& R125 + R134a + R143a  & \cite{vleR125R134aR161}    \\[-0.01cm]
${\rm CH_4}$ + ${\rm C_2H_4}$ + ${\rm C_2H_6}$  & \cite {vleCH4C2H4C2H6}	& R11 + R22 + R23 \dag			        & \cite{vleR23R22R114}     & R125 + R134a + R161 \dag  & \cite{vleR125R134aR161} \\[-0.01cm]
${\rm N_2}$ + ${\rm O_2}$ + ${\rm CO_2}$        & \cite {vleCO2O2N2}	& R12 + R113 + R152a \dag	       		  & \cite{vleR12R152aR113}   & R140a + R141b + R142b & \cite{vleR140aR142bR141b}  \\ \hline\hline
  					  						     
\end{tabular}				  						   
\end{center}				     						    
\end{table}										    
\clearpage


\listoffigures
\clearpage

\clearpage																																							

\clearpage																																							

\clearpage																																							

\clearpage																																							

\clearpage

\clearpage																																							

\clearpage																																							

\clearpage																																							

\clearpage																																							

\clearpage

\clearpage																																							

\clearpage																																							

\clearpage																																							

\clearpage																																							

\clearpage

\clearpage																																							

\clearpage																																							

\clearpage																																							

\clearpage																																							

\clearpage

\clearpage																																							

\clearpage																																							

\clearpage

\end{document}